\documentclass[aps,showpacs,amssymb,amsfonts,superscriptaddress,twocolumn,pra]{revtex4-1}

\usepackage{times}
\usepackage{bbm,mathrsfs}
\usepackage{graphicx}
\usepackage{amsfonts,amsmath}
\usepackage{amsthm}
\usepackage{color}
\usepackage[T1]{fontenc}
\usepackage{changes}

\def\textbf#1{{\bf #1}}
\def\be{\begin{equation}}
\def\ee{\end{equation}}
\def\ben{\begin{eqnarray}}
\def\een{\end{eqnarray}}
\def\eea{\end{array}}
\def\bea{\begin{array}}
\newcommand{\Tr}[0]{\mathrm{Tr}}
\newcommand{\ot}[0]{\otimes}
\newcommand{\bei}{\begin{itemize}}
\newcommand{\eei}{\end{itemize}}
\newcommand{\ket}[1]{|#1\rangle}
\newcommand{\bra}[1]{\langle#1|}

\newcommand{\proj}[1]{\ket{#1}\!\bra{#1}}

\usepackage{changes}

\definecolor{myurlcolor}{rgb}{0,0,0.7}
\definecolor{myrefcolor}{rgb}{0.8,0,0}
\usepackage{hyperref}
\hypersetup{colorlinks, linkcolor=myrefcolor,
citecolor=myurlcolor, urlcolor=myurlcolor}

\newtheorem{thm}{Theorem}

\newtheorem{cor}{Corollary}
\newtheorem{lemma}{Lemma}
\theoremstyle{definition}

\begin{document}

\title{Entangled symmetric states of $N$ qubits with all positive partial transpositions}

\author{R. Augusiak}\author{J. Tura}

\affiliation{ICFO--Institut de Ciencies Fotoniques, Mediterranean
Technology Park, 08860 Castelldefels (Barcelona), Spain}

\author{J. Samsonowicz}

\affiliation{Faculty of Mathematics and Information Science,
Warsaw University of Technology, Pl. Politechniki 1, 00-61
Warszawa, Poland}

\author{M. Lewenstein}

\affiliation{ICFO--Institut de Ciencies Fotoniques, Mediterranean
Technology Park, 08860 Castelldefels (Barcelona), Spain}
\affiliation{ICREA--Instituci\'o Catalana de Recerca i Estudis
Avan\c{c}ats, Lluis Companys 23, 08010 Barcelona, Spain}

\begin{abstract}
From both theoretical and experimental points of view symmetric
states constitute an important class of multipartite states.
Still, entanglement properties of these states, in particular
those with positive partial transposition (PPT), lack a systematic
study. Aiming at filling in this gap, we have recently
affirmatively answered the open question of existence of
four-qubit entangled symmetric states with positive partial
transposition and thoroughly characterized entanglement properties
of such states [J. Tura \textit{et al.}, \href{http://pra.aps.org/abstract/PRA/v85/i6/e060302}{Phys. Rev. A \textbf{85}, 060302(R) (2012)}]
With the
present contribution we continue on characterizing PPT entangled
symmetric states. On the one hand, we present all the results of
our previous work in a detailed way. On the other hand, we
generalize them to systems consisting of arbitrary number of
qubits. In particular, we provide criteria for separability of
such states formulated in terms of their ranks. Interestingly, for
most of the cases, the symmetric states are either separable or
typically separable. Then, edge states in these systems are
studied, showing in particular that to characterize generic PPT
entangled states with four and five qubits, it is enough to study
only those that assume few (respectively, two and three) specific
configurations of ranks. Finally, we numerically search for
extremal PPT entangled states in such systems consisting of up to
23 qubits. One can clearly notice regularity behind the ranks of
such extremal states, and, in particular, for systems composed of
odd number of qubits we find a single configuration of ranks for
which there are extremal states.
\end{abstract}


 \pacs{}

\maketitle

\section{Introduction}
\label{intro}

Characterization of entanglement \cite{HReview} in composite
quantum states with positive partial transposition (PPT states)
remains a difficult problem. One of the reasons for that is the
lack of a universal separability criterion allowing to distinguish
unambiguously separable from PPT entangled states (see,
nevertheless, e.g. Ref. \cite{GezaReview} for numerous necessary
separability conditions). There are, however, methods providing
some insight into the structure of PPT entangled states. One of
them exploits the fact that all states that remain positive under
partial transposition form a convex set, which as a proper subset
contains the PPT entangled states. To fully characterize the
latter, it is then enough to know all the extremal points of this
convex set. This approach has recently been extensively studied
(see Refs.
\cite{Leinaas,Leinaas2,Leinaas3,OptComm,ChenSkowronek}). In
particular, it allowed to solve the open problem of existence of
four-qubit PPT entangled symmetric states \cite{Tura}, and also,
although in an indirect way, disprove the Peres conjecture in the
multipartite case \cite{Tamas}.

The problem of characterization of PPT entangled states is even
more complicated in the multipartite case. Clearly, the set of PPT
states arises by intersecting sets of states that remain positive
under partial transpositions with respect to single bipartitions,
thus its boundary becomes more complicated with the increasing
number of parties. Nevertheless, the complexity can be reduced by
imposing some symmetries. For instance, demanding that the states
under study commute with multilateral action of unitary or
orthogonal groups leads to classes of multipartite states whose
full characterization with respect to entanglement becomes
possible (see e.g. Refs. \cite{groupstuff}).

Another interesting example of a class of states obtained by
imposing some symmetry are those supported on the symmetric
subspace of a given multipartite Hilbert space. The so-called
\textit{symmetric states} have recently been attracting much
attention
\cite{EckertAnP,Stockton,Ichikawa,Bastin,MathonetCenci,MeasSim,Perm1,Perm2}.
In particular, the underlying symmetry allowed for the use of the
Majorana representation \cite{Majorana} for an identification of
SLOCC classes of multipartite symmetric states \cite{Bastin} (see
also Refs. \cite{MathonetCenci}). The same symmetry provides
advantages in calculating certain entanglement measures
\cite{MeasSim}. Another motivation comes from the recent
experimental realizations of symmetric states of many-qubits, as
for instance, the six-qubit Dicke states \cite{DickeExp} or the
eight-qubit GHZ states \cite{GHZExp} (see also Ref. \cite{Bastin2}
in this context).

However, more effort has been devoted to the pure symmetric
states, leaving the characterization of entanglement of mixed, in
particular PPT states as an open problem. It is known so far that
for $N=2,3$ all PPT symmetric states are separable
\cite{EckertAnP}. Then, examples of five or six-qubit PPT
entangled symmetric states were found in Refs. \cite{Perm1,Perm2}.
Recently, the remaining case of $N=4$ has been studied in Ref.
\cite{Tura}, where the open question as to whether partial
transposition serves in this case as a necessary and sufficient condition for
separability (as this is the case for $N=2,3$) has been given
a negative answer.
The main aim of the present paper is to continue the
characterization of PPT entanglement in symmetric states. We
discuss in detail methods used in Ref. \cite{Tura} and then
generalize them to the case of arbitrary $N$. We derive
separability criteria for PPT symmetric states in terms of their
ranks and ranks of their partial transpositions. Then we exclude
configurations of ranks for which they are generically not edge.
Finally, we adapt to the multipartite case an algorithm allowing
to search for extremal PPT entangled states \cite{Leinaas} (see
also Ref. \cite{OptComm}). Exploiting it, we study ranks of the
extremal PPT entangled symmetric states consisting of up to $23$
qubits. Interestingly, we show that there are at most three
distinct configurations of ranks for which we find extremal PPT
entangled symmetric states, and, in particular, for odd $N$ there
is only a single such configuration.

The paper is structured as follows. In the next section (Sec.
\ref{prel}) we recall all notions and facts necessary for further
considerations. In Sec. \ref{sec:SymmStates} we investigate the
entanglement properties of the PPT symmetric states. Then, in Sec.
\ref{sec:Extr} we seek extremal entangled PPT symmetric states
consisting of even more than 20 qubits and classify them with
respect to their ranks. The results obtained for exemplary systems
consisting of four, five and six qubits are collected in Sec.
\ref{sec:special}. We conclude in Sec. \ref{sec:concl}.

\section{Preliminaries}
\label{prel}

Let us start with a couple of definitions that we will use
throughout the paper. Let
\begin{equation}
\mathcal{H}_{N}=\mathbbm{C}^{d_1}\ot\ldots\ot \mathbbm{C}^{d_N}
\end{equation}
denote a multipartite product Hilbert space and $D$ the convex set
of density operators acting on $\mathcal{H}_{N}$. By $R(\rho)$,
$r(\rho)$, $K(\rho)$, and $k(\rho)$ we denote, respectively, the
range, rank, kernel, and the dimension of the kernel of a given
$\rho\in D$. Then, $A_1,\ldots,A_N$ will stand for the subsystems
of a given $N$-partite $\rho$, and, in the case of low $N$, we
will also denote them by $A,B$, etc.

{\it PPT and separable states.} Let us now split the set
$I=\{A_1,\ldots,A_N\}$ into two disjoint subsets $S$ and
$\overline{S}$ $(S\cup\overline{S}=I)$ and call it bipartition
$S|\overline{S}$. We say that a given state $\rho$ acting on
$\mathcal{H}_{N}$ is PPT with respect to this bipartition iff
$\rho^{T_S}\geq0$. Clearly, states with this property make a
convex set denoted $D_S$. An element of $D$ whose partial
transpositions with respect to all bipartitions (notice that for a
given bipartition $S|\overline{S}$, partial transpositions with
respect to $S$ and $\overline{S}$ are equivalent under the full
transposition) are positive will be called {\it fully PPT}, and,
since in this paper we deal only with fully PPT states, we will be
calling them simply \textit{PPT states}. Clearly, such states make
also a convex set which is simply the intersection of $D_S$ for
all $S$.

A particular example of a state that is PPT is the \textit{fully
separable state} \cite{separable,DurCirac}:
\begin{equation}\label{separable}
\rho=\sum_{i}p_i\rho_{A_1}^{i}\ot\ldots\ot\rho_{A_{N}}^{i}, \quad
p_i\geq0,\quad \sum_ip_i=1,
\end{equation}
where $\rho_{A_j}^{i}$ denote density matrices representing all
subsystems. Clearly, in multipartite systems one may define
various types of separability (see e.g.
\cite{DurCirac,Acin3party}). Nonetheless, as we will see later, in
the symmetric case a given state $\rho$ is either genuine
multipartite entangled, i.e., cannot be written as a convex
combination of states which are separable with respect to in
general different bipartitions, or takes the form
(\ref{separable}).

{\it Edge states.} An important class of entangled PPT states are
the so-called {\it edge states} \cite{edge1,edge2,Karnas}. We call
$\rho$ acting on $\mathcal{H}_{N}$ edge iff there does not exist a
product vector $\ket{e_1}\ot\ldots\ot\ket{e_N}$ with
$\ket{e_i}\in\mathbbm{C}^{d_i}$ such that
$\ket{e_1}\ot\ldots\ot\ket{e_N}\in R(\rho)$ and
$(\ket{e_1}\ot\ldots\ot \ket{e_N})^{C_S}\in R(\rho^{T_S})$ for all
$S$, where by $C_S$ we denoted partial conjugation with respect to
$S$. The importance of edge states in the separability problem
comes from the fact that any PPT state can be decomposed as a
mixture of a fully separable and an edge state \cite{edge1}.
Alternatively speaking these are states from which no product
vector can be subtracted without losing the PPT or positivity
property, meaning that they lay on the boundary of the set of PPT
states. However, they do not have to be extremal, although any
extremal state is also edge.

Edge states have been studied in bipartite or three-partite
systems and many examples have been found
\cite{edge2,Kye1,Karnas}.

{\it Symmetric states.} Let us now concentrate on the $N$-qubit Hilbert space
\begin{equation}
\mathcal{H}_{2,N}=(\mathbbm{C}^2)^{\ot N}
\end{equation}
and consider its subspace $\mathcal{S}_N$ spanned by the unnormalized
vectors
\begin{equation}
\ket{E_i^N}=\ket{\{0,i\},\{1,N-i\}}\qquad (i=0,\ldots,N),
\end{equation}
which are just symmetric sums of vectors being products of $i$
zeros and $N-i$ ones. These vectors, when normalized, are also
known as Dicke states. For further benefits, let us notice that
the dimension of $\mathcal{S}_N$ is $N+1$, and therefore it is
isomorphic to $\mathbbm{C}^{N+1}$, which we denote
$\mathcal{S}_N\cong\mathbbm{C}^{N+1}$. Also, by $\mathcal{P}_N$ we
will be denoting the projector onto $\mathcal{S}_N$.

We call a state $\rho$ acting on $H_{2,N}$ {\it symmetric} iff it
is supported on $\mathcal{S}_N$, or, in other words, $R(\rho)\subseteq \mathcal{S}_N$.
In yet another words, $\rho$ is symmetric iff the equations
\begin{equation}
V_{\sigma}\rho=\rho V_{\sigma'}^{\dagger}=\rho
\end{equation}
are obeyed for any permutations $\sigma,\sigma'\in\Sigma_N$, where
$\Sigma_N$ is the group of all permutations of an $N$-element set,
while $V_{\sigma}$ is an operator defined as
$V_{\sigma}\ket{\psi_1}\ldots\ket{\psi_N}=\ket{\psi_{\sigma(1)}}
\ldots\ket{\psi_{\sigma(N)}}$
for any vectors $\ket{\psi_i}\in\mathbbm{C}^2$.

In the case of symmetric states the number of relevant partial
transpositions defining the set of PPT symmetric states
$D_{\mathrm{PPT}}^{\mathrm{sym}}$ is significantly reduced. This
is because positivity of a partial transposition with respect to
some subset $S$ is equivalent to positivity of all partial
transpositions with respect to subsystems of the same size $|S|$.
Together with the fact that for a given bipartition
$S|\overline{S}$, $\rho^{T_S}\geq
0\Leftrightarrow\rho^{T_{\overline{S}}}\geq 0$, one has $\lfloor
N/2\rfloor$ partial transpositions defining
$D_{\mathrm{PPT}}^{\mathrm{sym}}$. We choose them to be $T_{A_1}$,
$T_{A_1A_2}$, etc., however, for simplicity we will also be
denoting them as $T_1\equiv T_{A_1}$, $T_2\equiv T_{A_1A_2}$, and
so on. Alternatively, in systems of small size, we will use $T_A$,
$T_{AB}$, $T_{ABC}$, etc., to denote the relevant partial
transpositions.

Let us now notice that since
$\mathcal{S}_{i}\cong\mathbbm{C}^{i+1}$, a $N$-qubit symmetric
state $\rho$ can be seen with respect to a bipartition
$S|\overline{S}$ as a bipartite state acting on
$\mathbbm{C}^{|S|+1}\ot\mathbbm{C}^{N-|S|+1}$. This gives us
nontrivial bounds on the ranks of partial transpositions with
respect to all $S$, namely,
\begin{equation}
r(\rho^{T_S})\leq (|S|+1)(N-|S|+1)
\end{equation}
for $|S|=0,\ldots,\lfloor N/2\rfloor$, which, in particular means
that $r(\rho)\leq N+1$. A very convenient way of classifying PPT
states is through their ranks and ranks of their partial
transpositions, i.e., the $\lfloor N/2\rfloor$-tuples
\begin{eqnarray}\label{ranks}
&&\left(r(\rho),r(\rho^{T_{A_1}}),r(\rho^{T_{A_1A_2}}),\ldots, r(\rho^{T_{A_1\ldots A_{\lfloor N/2\rfloor}}})\right)\nonumber\\
&&\equiv\left(r(\rho),r(\rho^{T_1}),r(\rho^{T_2}),\ldots,r(\rho^{T_{\lfloor
N/2\rfloor}})\right).
\end{eqnarray}

Finally, let us recall that $\rho$ acting on some bipartite
Hilbert space
$\mathcal{H}_2=\mathbbm{C}^{d_1}\ot\mathbbm{C}^{d_2}$ is said to
be supported on $\mathcal{H}_2$ iff $R(\rho_A)=\mathbbm{C}^{d_1}$
and $R(\rho_B)=\mathbbm{C}^{d_2}$. Alternatively speaking, $\rho$
is not supported on $\mathcal{H}_2$ if either $\rho_A$ or $\rho_B$
has a vector in the kernel.

\section{Characterizing PPT entanglement in $N$-qubit symmetric states}
\label{sec:SymmStates}

Here, exploiting the results of Refs. \cite{2N,MN} we derive
separability criteria for the PPT symmetric states in terms of the
ranks (\ref{ranks}). Recall that in these papers it was shown that
any PPT state $\rho$ supported on a Hilbert space
$\mathbbm{C}^{d_1}\ot\mathbbm{C}^{d_2}$ is separable if
$r(\rho)\leq \max\{d_1,d_2\}$.

Then, we study the edge symmetric states and in particular we show
that symmetric states assuming certain ranks cannot be edge.

\subsection{Separability}
\label{sec:sep}

To begin with the separability properties let us recall that if a
pure $N$-partite symmetric state $\ket{\psi}$ is separable with
respect to some bipartition, then it must be fully separable
(\ref{separable}), i.e., $\ket{\psi}=\ket{e}^{\ot N}$ with
$\ket{e}\in\mathbbm{C}^2$ (see Refs. \cite{EckertAnP,Ichikawa}).
This straightforwardly implies that entangled symmetric pure and
thus mixed states have genuine multipartite entanglement (see also
Ref. \cite{Ichikawa}). Indeed, if a symmetric $\rho$ can be
written as a convex combination of density matrices, each
separable across some, in general different, bipartition, then
$\rho$ has pure separable vectors in its range. Since each such
vector is symmetric, it assumes the above form $\ket{e}^{\ot N}$,
meaning that $\rho$ is fully separable. Thus, throughout the
paper, by saying that a symmetric state $\rho$ is separable we
mean that it is fully separable, i.e., it takes the form
(\ref{separable}).

Let us now establish some conditions for separability in terms of
ranks of $\rho$. We start with the following technical lemma.

\begin{lemma}\label{lem_supp_v2}
Consider an $N$-qubit symmetric state and a bipartition
$S|\overline{S}$ $(|S|\leq N-|S|)$. Then, let $k_S$ and
$k_{\overline{S}}$ ($r_{S}$ and $r_{\overline{S}}$) denote the
dimensions of the kernels (ranges) of subsystems of $\rho$ with
respect to $S|\overline{S}$. The following statements hold:
\begin{itemize}
    \item if $k_{\overline{S}}>0$ then $r(\rho)\leq
    r_{\overline{S}}$,
    \item if $k_S>0$ then $r(\rho)\leq r_{S}$,
    \item if $k_{S}>0$ and $k_{\overline{S}}>0$, then $r(\rho)\leq
    \min\{r_S,r_{\overline{S}}\}$.
\end{itemize}
\end{lemma}
\begin{proof}
Recall first that with respect to the bipartition
$S|\overline{S}$, the $N$-qubit symmetric state $\rho$ can be seen
as a bipartite state acting on
$\mathbbm{C}^{|S|+1}\ot\mathbbm{C}^{|\overline{S}|+1}
=\mathbbm{C}^{|S|+1}\ot\mathbbm{C}^{N-|S|+1}$.

We will prove the first case and then the remaining two will
follow. Assume then that $k_{\overline{S}}>0$, meaning
that $\rho_{\overline{S}}$ has
$k_{\overline{S}}$ linearly independent vectors $\ket{\phi_i}$
$(i=1,\ldots,k_{\overline{S}})$ in the kernel. Consequently, for
any $|S|$-qubit symmetric vector $\ket{\psi}$, the projected
vectors $\mathcal{P}_N(\ket{\psi}\ket{\phi_i})$
$(i=1,\ldots,N-r_{\overline{S}}-|S|+1)$ belong to $K(\rho)$. In
what follows we will show that by choosing properly vectors
$\ket{\psi}$, one is able to find $N-r_{\overline{S}}+1$ linearly
independent vectors in $K(\rho)$ of this form.

First, we prove that for any $\ket{\psi}$,
$N-r_{\overline{S}}-|S|+1$ projected vectors
$\mathcal{P}_N(\ket{\psi}\ket{\phi_i})\in K(\rho)$ are linearly
independent. Towards this end, let us assume, in contrary, that
there exists a collection of nonzero numbers
$\alpha_i\in\mathbbm{C}$ such that
$\sum_i\alpha_i\mathcal{P}_N(\ket{\psi}\ket{\phi_i})=0$. The
latter is equivalent to saying that the vector $\ket{\psi}\ot
\sum_i\alpha_i\ket{\phi_i}$ sits in the kernel of $\mathcal{P}_N$.
But since $\sum_i\alpha_i\ket{\phi_i}$ is an $|S|$-qubit symmetric
vector, this is possible only if $\ket{\phi_i}$ are linearly
dependent, contradicting the fact that they span the kernel of
$\Tr_S\rho$.

Now, we consider particular vectors
\begin{equation}\label{Phi}
\ket{\Phi_i^j}=\mathcal{P}_N(\ket{E^{|S|}_j}\ket{\phi_i})
\end{equation}
with $j=0,\ldots,|S|$ and
$i=1,\ldots,|\overline{S}|-r_{\overline{S}}+1$. As already proven,
for any $j$, the vectors $\ket{\Phi_i^{j}}$ make
an ($|\overline{S}|-r_{\overline{S}}+1$)-element linearly independent set.
Let us now concentrate on the vectors $\ket{\Phi_i^{0}}$ and
choose the one for which $\langle
E_{|\overline{S}|}^{N}|\Phi_i^{0}\rangle$ is nonzero, say
$\ket{\Phi_l^{0}}$. Notice that by the construction $\langle
E_{k}^{N}|\Phi_i^{0}\rangle=0$ for $k>|\overline{S}|$; to obtain
$\ket{\Phi_i^{0}}$ we symmetrize $\ket{\phi_i}$ with
$\ket{E^{|S|}_0}=\ket{0}^{\ot |S|}$, meaning that
$\ket{\Phi_i^{0}}$ decomposes into the symmetric vectors
$\ket{E^N_{k}}$ with $k\leq |\overline{S}|$.
If, however, $\langle E_{|\overline{S}|}^{N}|\Phi_i^{0}\rangle=0$ for all $i$, we choose the one
for which $\langle E_{|\overline{S}|-1}^{N}|\Phi_i^{0}\rangle\neq 0$, etc.
Clearly, repeating this we must find the desired vector as otherwise $\ket{\Phi_i^0}=0$ for all $i$, contradicting
the fact that the vectors $\ket{\Phi_i^0}$ are linearly independent.

Let us then assume for simplicity that $\ket{\Phi_l^{0}}$ is such
that $\langle E_{|\overline{S}|}^{N}|\Phi_l^{0}\rangle\neq 0$,
meaning that $\bra{\phi_l}1\rangle^{\ot|\overline{S}|}\neq 0$ [cf.
Eq. (\ref{Phi})]. Consequently, the vectors $\ket{\Phi_l^j}$ $(j=1,\ldots,|S|)$,
when decomposed into the symmetric basis of $\mathcal{S}_N$,
contain $\ket{E^N_k}$ with $k\geq |\overline{S}|+1$ and therefore
are linearly independent of the set $\{\ket{\Phi_i^0}\}_i$.
Moreover, by the very construction, they make an $|S|$-element set
of linearly independent vectors themselves, meaning that
the vectors $\ket{\Phi_i^0}$
$(i=1,\ldots,|\overline{S}|-r_{\overline{S}}+1)$ together with
$\ket{\Phi_l^j}$ $(j=1,\ldots,|S|)$ make the desired set of
$N-r_{\overline{S}}+1$ linearly independent vectors in $K(\rho)$.

Consequently, $k(\rho)\geq N-r_{\overline{S}}+1$
which, taking into account the maximal possible rank of a
symmetric state $\rho$, gives the bound $r(\rho)\leq
N+1-N+r_{\overline{S}}-1=r_{\overline{S}}$. In an analogous way
one proves the second case, i.e., when $k_S>0$. Precisely, following the above arguments,
one sees that $k_S$ linearly independent vectors in the kernel of
$\rho_S$ gives at least $N-r_S+1$ linearly independent vectors in
$K(\rho)$, imposing the bound $r(\rho)\leq r_S$. To prove the
third case, one just chooses the tighter of both the above bounds,
i.e., $r(\rho)\leq \min\{r_S,r_{\overline{S}}\}$.
\end{proof}

Essentially, this lemma says that if the symmetric state $\rho$ is
not supported on $\mathbbm{C}^{|S|+1}\ot\mathbbm{C}^{N-|S|+1}$
with respect to the bipartition $S|\overline{S}$, its rank is
bounded from above by ranks of its subsystems. In the particular
case when the subsystem $S$ consists of a single party, it
straightforwardly implies that if $r(\rho)\geq N$ then $\rho$ has
to be supported on $\mathbbm{C}^2\ot\mathbbm{C}^N$ with respect to
the bipartition one versus the rest ($A_1|A_2\ldots A_N$).

The following fact was already stated in Ref. \cite{EckertAnP},
however, a detailed proof was not given. We will exploit lemma
\ref{lem_supp_v2} to demonstrate it rigorously.

\begin{thm}\label{thm_sep1}
Let $\rho$ be a $N$-qubit PPT symmetric state. If it is entangled
then $r(\rho)=N+1$, i.e., $\rho$ is of maximal rank.
\end{thm}
\begin{proof}An $N$-qubit symmetric state can be seen
as a bipartite state acting on $\mathbbm{C}^2\ot\mathbbm{C}^N$
with respect to the bipartition one qubit versus the rest, as for
instance $A_1|A_2\ldots A_N$. Let us denote by $\rho_{A_1}$ and
$\rho_{A_1\ldots A_N}$ the subsystems of $\rho$ with respect to
this bipartition. Assuming then that $r(\rho)\leq N$, the results
of Ref. \cite{2N} imply that $\rho$ is separable provided it is
supported on $\mathbbm{C}^2\ot\mathbbm{C}^N$. If, however, the
latter does not hold, there are vectors in the kernel of either
$\rho_{A_1}$ or $\rho_{A_2\ldots A_N}$. In the first case, lemma
\ref{lem_supp_v2} implies that $r(\rho)=1$ and $\rho$ is a pure
product vector, while in the second case $r(\rho)$ is upped
bounded by the rank of $\rho_{A_2\ldots A_N}$. Again, results of
Ref. \cite{2N} apply here, meaning that $\rho$ is separable.
\end{proof}

Alternatively speaking, this theorem means that there are no PPT
entangled $N$-qubit symmetric states of rank less than $N+1$. On
the other hand, it provides only a sufficient condition for
separability, as there are separable symmetric states of rank
$N+1$. Another consequence of theorem \ref{thm_sep1} 
is that for an arbitrary bipartition $S|\overline{S}$, a PPT
entangled state $\rho$ and its partial transposition $\rho^{T_S}$
are supported on the corresponding Hilbert space
$\mathbbm{C}^{|S|+1}\ot \mathbbm{C}^{N-|S|+1}$. Specifically, one
has the following lemma.

\begin{lemma}\label{cor2_sep}
Consider an $N$-qubit PPT symmetric state and an arbitrary bipartition
$S|\overline{S}$. If $\rho$ is entangled, then $\rho^{T_S}$
is supported on the bipartite Hilbert space corresponding to
the bipartition $S|\overline{S}$, i.e., $\mathbbm{C}^{|S|+1}\ot
\mathbbm{C}^{N-|S|+1}$. In other words, if for some bipartition $S|\overline{S}$,
$\rho^{T_S}$ is not supported on $\mathbbm{C}^{|S|+1}\ot
\mathbbm{C}^{N-|S|+1}$, then $\rho$ is separable.
\end{lemma}
\begin{proof}
Assume that the PPT state $\rho$ is entangled but the partial
transposition $\rho^{T_S}$ is not supported on the Hilbert space
corresponding to the bipartition $S|\overline{S}$, i.e.,
$\mathbbm{C}^{|S|+1}\ot \mathbbm{C}^{N-|S|+1}$. This means that
one of its subsystems, say the $\overline{S}$ one, contains an
$|\overline{S}|$-qubit symmetric vector $\ket{\phi}$ in the
kernel. Consequently, for any $|S|$-qubit vector $\ket{\psi}$, the
following implication
\begin{equation}
\rho^{T_S}\ket{\psi}\ket{\phi}=0 \quad\Rightarrow\quad
\rho\ket{\psi^{*}}\ket{\phi}=0
\end{equation}
holds. Putting for instance $\ket{\psi}=\ket{0}^{\ot |S|}$, one
sees that the symmetrized vector $\mathcal{P}_N(\ket{0}^{\ot
|S|}\ot\ket{\phi})$ belongs to $K(\rho)$. As a result $r(\rho)\leq
N$ and theorem \ref{thm_sep1} implies that $\rho$ is separable,
leading to the contradiction.
\end{proof}

Then, with the aid of lemma \ref{cor2_sep}, we can prove
the analog of theorem \ref{thm_sep1} for the ranks of partial transpositions of
$\rho$.

\begin{thm}\label{thm_sep2}
Let us consider an $N$-qubit PPT symmetric state $\rho$ and a
bipartition $S|\overline{S}$ with an arbitrary $S$ $(|S|\leq
|\overline{S}|)$. If it is entangled then $r(\rho^{T_S})>
N-|S|+1$. In particular, if $\rho$ is entangled then
$r(\rho^{T_A})\geq  N+1$.
\end{thm}
\begin{proof}Due to lemma \ref{cor2_sep}, we can assume that
with respect to the bipartition on $S|\overline{S}$, $\rho$ is
supported on $\mathbbm{C}^{|S|+1}\ot \mathbbm{C}^{N-|S|+1}$ as
otherwise it is separable. Then, if $r(\rho^{T_S})\leq N-|S|+1$,
the results of Ref. \cite{MN} imply that $\rho$ is separable.
Noting that this reasoning is independent of the bipartition, we
complete the proof.
\end{proof}

In other words, any PPT symmetric states whose rank obeys
$r(\rho^{T_S})\leq N-|S|+1$ for some bipartition, is separable.

Still, by using more tricky bipartitions we can provide further
separability conditions for generic symmetric states in terms of
the ranks. In this direction we prove the following theorem.

\begin{thm}\label{sep_gen}
Consider an $N$-qubit PPT symmetric state $\rho$ and a bipartition
$S|\overline{S}$. If $r(\rho^{T_S})\leq(|S|+1) (N-|S|)$, then the
generic $\rho$ is separable.
\end{thm}
\begin{proof}
Consider an $N$-qubit state $\sigma=\rho^{T_S}$ which is PPT, but
no longer symmetric, and a party which does not belong to $S$, say
$A_N$. With respect to the bipartition $A_N$ versus the rest,
$\sigma$ can be seen as a bipartite state acting on
$\mathbbm{C}^{(|S|+1)(N-|S|)}\ot \mathbbm{C}^2$. Since $\rho$ is
fully PPT, it clearly follows that $\sigma^{T_{A_N}}\geq 0$. This,
together with the fact that $r(\rho^{T_S})\leq (|S|+1)(N-|S|)$,
implies that $\sigma$ has to be separable across the bipartition
$A_1\ldots A_{N-1}|A_N$, i.e.,
\begin{equation}\label{dupa0}
\sigma=\rho^{T_S}=\sum_{i}\proj{\psi_i}_{A_1\ldots A_{N-1}}\ot \proj{e_i}_{A_N},
\end{equation}
provided that it is supported on $\mathbbm{C}^{(|S|+1)(N-|S|)}\ot
\mathbbm{C}^2$, which generically is the case.

Now, let us notice that $\rho^{T_{S}}$ is still symmetric with
respect to subsystem $\overline{S}$, i.e.,
$\mathcal{P}_{\overline{S}}\rho^{T_{S}}\mathcal{P}_{\overline{S}}=\rho^{T_{S}}$.
Since then all vectors $\ket{\psi_i}\ket{e_i}$ appearing in the
decomposition (\ref{dupa0}) belong to $R(\rho^{T_{S}})$, they also
enjoy the above symmetry, i.e.,
$\mathcal{P}_{\overline{S}}\ket{e_i}_{A_N}\ket{\psi_i}_{A_1\ldots
A_{N-1}}=\ket{e_i}_{A_N}\ket{\psi_i}_{A_1\ldots A_{N-1}}$ for all
$i$. This, as it is shown below, implies that for any $i$,
$\ket{\psi_i}=\ket{\widetilde{\psi}_{i}}\ket{e_i}^{\ot
|\overline{S}|}$ with $\ket{\widetilde{\psi}_i}$ being from the
Hilbert space corresponding to the subsystem $S$. Putting these
forms to Eq. (\ref{dupa0}), one arrives at
\begin{equation}
\rho^{T_{S}}=\sum_{i}\proj{\widetilde{\psi}_i}_{S}\ot
\proj{e_i}^{\ot |\overline{S}|}.
\end{equation}
Now, one can move the partial transposition with respect to $S$ to
the right-hand side of the above identity, and use the fact that
$\rho$ is symmetric, which leads us to the form (\ref{separable}).

To complete the proof let us show that if $\mathcal{P}_{A\cup
S}\ket{e}_A\ket{\psi}_{\overline{S}\cup
S}=\ket{e}_A\ket{\psi}_{\overline{S}\cup S}$ holds for some
one-qubit and $N$-qubit vectors $\ket{e}$ and $\ket{\psi}$, then
the latter assumes the form $\ket{\psi}_{\overline{S}\cup
S}=\ket{\widetilde{\psi}}_{\overline{S}}\ot\ket{e}^{\ot |S|}$ with
$\ket{\widetilde{\psi}}$ belonging to the Hilbert space associated
to the subsystem $\overline{S}$. Here by $\mathcal{P}_{A\cup S}$
we denote a projector onto the symmetric subspace of the Hilbert
space corresponding to the qubits $A$ and $S$.

To this end, let us decompose
\begin{equation}\label{yet}
\ket{\psi}=\sum_i\ket{i}_{\overline{S}}\ket{\phi_i}_{S},
\end{equation}
where $\{\ket{i}\}$ denotes any orthogonal basis in the Hilbert
space associated to the subsystem $\overline{S}$. It is clear that
$\mathcal{P}_{A\cup S}\ket{e}_A\ket{\psi}_{\overline{S}\cup
S}=\ket{e}_A\ket{\psi}_{\overline{S}\cup S}$ is equivalent to
$\mathcal{P}_{A\cup
S}\ket{e}_A\ket{\phi_i}_S=\ket{e}_A\ket{\phi_i}_S$ for all $i$. By
virtue of the results of Refs. \cite{EckertAnP,Ichikawa}, the
latter can hold only if $\ket{\phi_i}=\ket{e}^{\ot |S|}$ for any
$i$. For completeness, let us recall the proof of this fact. For
this purpose, assume that
$\ket{e}\ket{\phi}\in\mathcal{S}_{|S|+1}$ for some one-qubit and
$|S|$-qubit vectors $\ket{e}$ and $\ket{\phi}$, respectively.
Then, one immediately concludes that $\ket{\phi}$ must be
symmetric and consequently can be written as
\begin{equation}
\ket{\phi}=\sum_{i=1}^{|S|+1}\alpha_i\ket{E_i^{|S|}}
\end{equation}
with $\alpha_i\in\mathbbm{C}$. Putting $\ket{e}=(a,b)$ and utilizing the fact that
$\mathcal{P}_{|S|+1}\ket{e}\ket{\phi}=\ket{e}\ket{\phi}$, one has
\begin{eqnarray}
&&\sum_{i=1}^{|S|+1}\alpha_i\left[\frac{\binom{|S|}{i-1}}
{\binom{|S|+1}{i-1}}\ket{E_i^{|S|+1}}+\frac{\binom{|S|}{i-1}}
{\binom{|S|+1}{i}}\ket{E_{i+1}^{|S|+1}}\right]\nonumber\\
&&=(a\ket{0}+b\ket{1})\ot
\sum_{i=1}^{|S|+1}\alpha_i\ket{E_i^{|S|}}.
\end{eqnarray}
Projection of the above onto vectors $\ket{0}^{\ot (|S|+1)}$,
$\ket{0\ldots 01}$, $\ket{0\ldots 011}$, etc., leads to equations
$a\alpha_{j}=b\alpha_{j-1}$ $(j=2,\ldots,|S|+1)$, which, in turn,
imply that $\alpha_j=(b/a)^{j-1}\alpha_1$ $(j=2,\ldots,|S|+1)$.

Putting now $\ket{\phi_i}=\ket{e}^{\ot |S|}$ for all $i$ to Eq. (\ref{yet}), one gets
\begin{equation}
\ket{\psi}=\sum_{i}\ket{i}_{\overline{S}}\ket{\psi_i}_S=\sum_{i}\ket{i}_{\overline{S}}\ket{e}^{\ot
|S|}=\ket{\widetilde{\psi}}_{\overline{S}}\ket{e}^{\ot |S|},
\end{equation}
which finishes the proof.
\end{proof}

Notice that the above theorems imply that for most of the possible
ranks, the symmetric states are either separable or generically
separable. The maximal rank of a partial transposition with
respect to a given subsystem $S$ is $(|S|+1)(N-|S|+1)$. Taking
into account the fact that we can put $r(\rho)=N+1$ (as otherwise
the state is separable, cf. theorem \ref{thm_sep1}), we have in total
\begin{equation}\label{total}
\prod_{|S|=1}^{\lfloor N/2\rfloor}(|S|+1)(N-|S|+1)=
N!(N/2+1)^{2\left(\lfloor N/2 \rfloor -N/2  \right)+1}
\end{equation}
possible configurations of the relevant ranks (\ref{ranks}) that
can be assumed by the symmetric states. With theorems 
\ref{thm_sep2} and \ref{sep_gen}, we see that for a transposition
with respect to $S$, symmetric states of the first
$(|S|+1)(N-|S|)$ of the corresponding ranks are either separable
or generically separable. This leaves only $|S|+1$ of ranks with
respect to this bipartition for which they do not have to be
generically separable. Taking into account all the relevant
partial transpositions, we have in total $(\lfloor N/2\rfloor+1)!$
of the remaining cases where one can search for PPT entangled
symmetric states. This is clearly a small portion (rapidly
vanishing for large $N$) of all the possible ranks [cf. Eq.
(\ref{total})]. For instance, for $N=4,5$, this gives us treatable
$6$ configurations of ranks (out of all, respectively, $72$ and
$120$ obtained from Eq. (\ref{total})), while for $N=6,7$ this
number amounts to $24$ (Eq. (\ref{total}) gives in these cases,
respectively, $2880$ and $5040$ possible ranks). As we will see
later, if additionally we ask about possibility of being an edge
state, these numbers may be further reduced.

\subsection{$N$-qubit symmetric edge states}
\label{sec:edge}

Let us now single out the configurations of ranks where the
symmetric states can be edge. Clearly, theorem \ref{sep_gen}
implies that the ranks of $\rho^{T_S}$ for all $S$ must be larger
than $(|S|+1)(N-|S|)$ as otherwise the generic symmetric states
are separable. In what follows we will provide a few results
allowing to bound the ranks from above.

In general, as already discussed in Sec. \ref{prel}, to prove that
a given $\rho$ is not edge one has to prove that there is
$\ket{e}\in\mathbbm{C}^2$ such that $\ket{e}^{\ot N}\in R(\rho)$
and $[\ket{e}^{\ot N}]^{C_S}\in R(\rho^{T_S})$ for all $S$.
Assuming that the rank of $\rho$ is maximal, $r(\rho)=N+1$, the
above is equivalent to solving of a system of
$\sum_{|S|=1}^{\lfloor N/2\rfloor}k(\rho^{T_S})$ equations
\begin{equation}\label{Eqs}
[\bra{e}^{\ot N}]^{C_S}\ket{\Psi_i^{S}}=0,
\end{equation}
where $\ket{\Psi_i^{S}}\in K(\rho^{T_S})$ and $S=A,AB,\ldots$. By
putting $\ket{e}=(1,\alpha)$, one reduces Eqs. (\ref{Eqs}) to a
system of polynomial equations $P(\alpha,\alpha^{*})=0$ in
$\alpha$ and $\alpha^{*}$. This is clearly a hard problem to solve
(see, e.g., the discussion in Ref. \cite{2N}). Still, under some
assumptions and using a method of Ref. \cite{Normal}, it is
possible to find a solution to a single equation of that type.

\begin{lemma}\label{lem_pol}
Consider an equation
\begin{equation}\label{eq}
\sum_{i=0}^{k}(\alpha^{*})^{i}Q_{i}(\alpha)=0\qquad
(\alpha\in\mathbbm{C}),
\end{equation}
where $Q_i$ $(i=0,\ldots,k)$ are some polynomials. If
$\max_i\{\deg Q_i\}=\deg Q_k= n>k$ and $\deg Q_0=m>k$, then this
equation has at least one solution.
\end{lemma}
\begin{proof}
Notice that, via the results of Ref. \cite{2N}, Eq. (\ref{eq}) has
generically at most $2^{k-1}[k+n(n-k+1)]$ complex solutions. To
find one, in Eq. (\ref{eq}) we substitute $\alpha=rs$ and
$\alpha^{*}=r/s$ with $r\in\mathbbm{R}$ and $s\in\mathbbm{C}$, obtaining
\begin{equation}\label{eqNowe}
\sum_{i=0}^ks^{k-i}r^{i}Q_{i}(rs)=0.
\end{equation}
Treating $r$ as a parameter and $s$ as a variable, our aim now is
to prove that for some $r$ there is $s$ such that $|s|=1$ and Eqs.
(\ref{eq}) and (\ref{eqNowe}) is obeyed. For this purpose, let us first put $s=x/r$
with $x\in\mathbbm{C}$, which gives us
\begin{equation}\label{Eq2}
\sum_{i=0}^{k}x^{k-i}r^{2i}Q_{i}(x)=0.
\end{equation}
In the limit $r\to\infty$, the left-hand side of the above
equation approaches $Q_k(x)$, meaning that Eq. (\ref{eqNowe}) has $n$
solutions $s^{\infty}_{i}\to 0$ $(i=1,\ldots,n)$. Then, in the
limit of $r\to 0$, the left-hand side of Eq. (\ref{Eq2}) goes to
$Q_0(\alpha)$, implying that Eq. (\ref{eqNowe}) has $m$ solutions
$s^{0}_{i}\to\infty$ $(i=1,\ldots,m)$.

Then, one sees that Eq. (\ref{eqNowe}) has at most $n+k$ solutions
with respect to $s$. Consequently, for $r\to \infty$ and $r\to 0$,
Eq. (\ref{eqNowe}) has additional $k$ roots $s_i^{\infty}$
$(i=n+1,\ldots,n+k)$ and $n+k-m$ roots $s_{i}^{0}$ $(i=m+1,\ldots,n+k)$,
respectively, which can remain unspecified. As $r$ varies continuously from
zero to large values, all roots $s_i^{0}$ must continuously
tend to $s_i^{\infty}$. But, since $m>k$, at
least one of $m$ roots $s_i^{0}\to\infty$ $(i=1,\ldots,m)$
must tend to one of the $n$ roots $s_i^{\infty}$
$(i=1,\ldots,n)$ which are close
to zero. 
This means that there is at least one pair $(r,s)$ with $|s|=1$
solving Eq. (\ref{eqNowe}) and thus Eq, (\ref{eq}).
\end{proof}

With the aid of the above lemma we can prove the following
theorem.

\begin{thm}\label{thm:edge0}
Let us consider $N$-qubit PPT symmetric state and a subsystem $S$
of size $1\leq|S|\leq \lceil N/2\rceil-1$, and assume that
$r(\rho^{T_X})=(|X|+1)(N-|X|+1)$ (maximal) for all $X$ except for
$X=S$ for which $r(\rho^{T_S})=(|S|+1)(N-|S|+1)-1$. Then,
generically such states are not edge.
\end{thm}
\begin{proof}We prove that under the above assumptions it is
generically possible to find a product vector $\ket{e}^{\ot N}\in
R(\rho)$ such that $(\ket{e}^{\ot N})^{C_X}\in R(\rho^{T_X})$ for
all subsystems $X$. Clearly, all ranks of $\rho$ are maximal
except for the one corresponding to the partial transposition with
respect to the subsystem $S$, which is
$r(\rho^{T_S})=(|S|+1)(N-|S|+1)-1$ (maximal diminished by one).
Denoting by $\ket{\Psi}$ the unique vector from the kernel of
$\rho^{T_S}$, one then has to solve a single equation
$\bra{\Psi}(\ket{e^{*}}^{\ot |S|}\ot \ket{e}^{\ot (N-|S|)})=0$.
After putting $\ket{e}=(1,\alpha)$ with $\alpha\in\mathbbm{C}$,
the latter can be rewritten as
\begin{equation}\label{pol_eq}
\sum_{i=0}^{|S|}(\alpha^{*})^{i}Q_{i}(\alpha)=0,
\end{equation}
where $Q_i(\alpha)$ $(i=0,\ldots,|S|)$ are polynomials of degree
at most $N-|S|$ and generically they are exactly of degree
$N-|S|$. Due to the assumption that $N-|S|>|S|$, lemma
\ref{lem_pol} applies here, implying that (\ref{pol_eq}) has at
least one solution and generic symmetric $\rho$ of the above ranks
is not edge.
\end{proof}

The above theorem says that generic PPT symmetric states having
all ranks maximal except for a single one corresponding to a
partial transposition with respect to a subsystem $S$ such that
$|S|<N-|S|$, for which the rank is one but maximal, are not edge.
This method, however, does not work for even $N$ and in the case
when the chosen partial transposition is taken with respect to the
half of the whole system, i.e., $|S|=|\overline{S}|=N-|S|$. This
is because in this case the resulting equation (\ref{Eq2}) is of
the same orders in $\alpha$ and $\alpha^{*}$ and the above method
does not apply.

Still, however, using a different approach, we can prove an
analogous fact for $N=4$ and $N=6$. Specifically, we will show
that all symmetric states of all ranks maximal except for the last
one, which is one less than maximal, are not edge. Towards this
end, let us start with some general considerations.

Let $\rho$ be an $N$-qubit symmetric state with even $N$, and let
now $S$ denote a particular bipartition consisting of first $N/2$
qubits (half of the state). Assume then that all ranks of $\rho$
are maximal except for the one corresponding to the partial
transposition with respect to $S$, for which it is one but
maximal, i.e., $(N/2+1)^2-1$. Consequently, there is a single
vector $\ket{\Psi}$ in the kernel of $\rho^{T_S}$, meaning that
$\rho$ is not edge iff a single equation [cf. Eqs. (\ref{Eqs})]:
\begin{equation}\label{rownanie}
\bra{e}^{\ot N/2}\bra{e^{*}}^{\ot N/2} \ket{\Psi}=0
\end{equation}
with $\ket{e}\in\mathbbm{C}^2$ has a solution. For this purpose,
one notices that
\begin{equation}
V_{S,\overline{S}}\rho^{T_S}V^{\dagger}_{S,\overline{S}}=(\rho^{*})^{T_S},
\end{equation}
with $V_{S,\overline{S}}$ denoting a unitary operator swapping the
subsystems $S$ and $\overline{S}$ (notice that both are of the
same size). Then, because $\ket{\Psi}$ is the unique vector in
$K(\rho^{T_S})$, it has to enjoy the same symmetry, i.e.,
$V_{S,\overline{S}}\ket{\Psi}=\ket{\Psi^{*}}$. Denoting now by
$M_{\Psi}$ the matrix representing elements of $\ket{\Psi}$ in the
product basis $\ket{E^{N/2}_i}\ket{E^{N/2}_j}$ of
$\mathcal{S}_{N/2}\ot \mathcal{S}_{N/2}$, the latter symmetry
implies that $M_{\Psi}=M_{\Psi}^{\dagger}$. Diagonalizing
$M_{\Psi}$, one sees that $\ket{\Psi}$ can be written in the
following form
\begin{equation}\label{form}
\ket{\Psi}=\sum_{l=1}^{N/2+1}\lambda_l\ket{\omega_l^{*}}\ket{\omega_l}
\end{equation}
with $\lambda_l\in\mathbbm{R}$ and $\ket{\omega_l}\in S_{N/2}$
being eigenvalues and eigenvectors of $M_{\Psi}$, respectively.

The fact that $\ket{\Psi}\in K(\rho^{T_S})$ implies that for any
pair $\ket{x},\ket{y}\in\mathcal{S}_{N/2}$, one has
\begin{eqnarray}
\bra{x^{*},y}\rho^{T_S}\ket{\Psi}&=&\sum_l\lambda_l\bra{x^{*},y}\rho^{T_S}\ket{\omega_l^{*},\omega_l}\nonumber\\
&=&\sum_l\lambda_l\bra{\omega_l,y}\rho\ket{x,\omega_l}\nonumber\\
&=&\sum_l\lambda_l\bra{\omega_l,y}\rho\ket{\omega_l,x}=0,
\end{eqnarray}
where the third equality follows from the fact that
$V_{S,\overline{S}}\rho=\rho$. As a result
\begin{equation}\label{dziwne}
\Tr\left[(W\ot\ket{x}\!\bra{y})\rho\right]=0
\end{equation}
holds for any pair of vectors
$\ket{x},\ket{y}\in\mathcal{S}_{N/2}$, where
$W=\sum_l\lambda_l\proj{\omega_l}$.

On the other hand, with the aid of Eq. (\ref{form}), one can
rewrite Eq. (\ref{rownanie}) as
\begin{equation}\label{next_eq}
\bra{e}^{\ot N/2}W\ket{e}^{\ot N/2}=0.
\end{equation}
Assume, in contrary, that the above (equivalently Eq.
(\ref{rownanie})) does not have any solution. Then, its left-hand
side must have the same sign for any $\ket{e}\in\mathbbm{C}^2$,
say positive. This means that the matrix $W$ is an $N/2$-qubit
entanglement witness supported on the symmetric subspace
$\mathcal{S}_{N/2}$. We already know that there are no PPT
entangled symmetric states of two and three qubits, and hence for
$N=4$ and $N=6$, this witness must be decomposable. Precisely
\begin{equation}\label{witness2}
W=P+Q^{T_A}
\end{equation}
for $N=4$ with $P,Q$ being positive matrices acting on
$\mathbbm{C}^2\ot\mathbbm{C}^2$, while
\begin{equation}\label{witness3}
W=\widetilde{P}+\widetilde{Q}^{T_A}+\widetilde{R}^{T_{AB}}
\end{equation}
for $N=6$ with $\widetilde{P},\widetilde{Q},\widetilde{R}\geq 0$
acting on $\mathbbm{C}^{2}\ot\mathbbm{C}^2\ot\mathbbm{C}^2$.
Putting Eqs. (\ref{witness2}) and (\ref{witness3}) to Eq.
(\ref{dziwne}), one in particular arrives at the following
conditions
\begin{equation}
\Tr\left[\left(P\ot\proj{x}\right)\rho\right]=
\Tr\left[\left(Q\ot\proj{x}\right)\rho^{T_A}\right]=0.
\end{equation}
for $N=4$ and
\begin{eqnarray}
\Tr[(\widetilde{P}\ot\proj{x})\rho]
&=&\Tr[(\widetilde{Q}\ot\proj{x})\rho^{T_A}]\nonumber\\
&=&\Tr[(\widetilde{R}\ot\proj{x})\rho^{T_{AB}}]=0
\end{eqnarray}
for $N=6$ and for any $\ket{x}\in\mathcal{S}_{N/2}$. These
conditions imply that either $\rho$ or $\rho^{T_A}$, or
$\rho^{T_{AB}}$ is not of full rank, contradicting the assumption.
Therefore, Eq. (\ref{next_eq}) and thus Eq. (\ref{rownanie}) must
have a solution. In this way we have proven the following theorem.

\begin{thm}\label{thm:edge1}
Four-qubit symmetric states of ranks $(5,8,8)$ and six-qubit
symmetric states of ranks $(7,12,15,15)$ are not edge.
\end{thm}

We already know that for there exist PPT entangled symmetric
states consisting of more than three qubits
\cite{Perm1,Perm2,Tura} and thus indecomposable entanglement
witnesses detecting them. Consequently, the above method does not
apply in general for even $N\geq 8$.
Nevertheless, it provides a necessary condition for being edge,
i.e., if a symmetric state of all ranks maximal except for
$r(\rho^{T_{N/2}})$ which is one less than maximal is edge, the
witness $W=\sum_l\lambda_l\proj{\omega_l}$ constructed from
(\ref{form}) is indecomposable.

Interestingly, for $N=4$ we can prove analogous theorem in the
case when the rank of $r(\rho^{T_{AB}})$ is two less than maximal.

\begin{thm}\label{thm:edge2}
Generic four-qubit symmetric states of ranks $(5,8,7)$ are not
edge.
\end{thm}
\begin{proof}By assumption, $\rho$ and $\rho^{T_A}$ are of full rank, while
$K(\rho^{T_{AB}})$ contains two linearly independent vectors
$\ket{\Psi_i}$ $(i=1,2)$. Consequently, to find a product vector
$\ket{e}^{\ot 4}\in R(\rho)$ such that $\ket{e^*}^{\ot
2}\ket{e}^{\ot 2}\in R(\rho^{T_{AB}})$, one has to solve two
equations
\begin{equation}
\bra{e^*}^{\ot 2}\bra{e}^{\ot 2}|\Psi_i\rangle=0\qquad(i=1,2).
\end{equation}

Let us now briefly characterize the vectors $\ket{\Psi_i}$. With
the aid of the identity
$\rho^{T_{AB}}=V_{AB,CD}(\rho^{*})^{T_{AB}}V_{AB,CD}$, where
$V_{AB,CD}$ is a unitary operator swapping $AB$ and $CD$
subsystems, one may show that they can be written as
\begin{equation}\label{wektory}
\ket{\Psi_1}=\sum_{k=1}^{2}\lambda_{k}\ket{e_k}\ket{f_k^{*}},\qquad
\ket{\Psi_2}=\sum_{k=1}^{2}\lambda_{k}\ket{f_k}\ket{e_k^{*}}.
\end{equation}
Indeed, let us first notice that we can assume that one of
$\ket{\Psi_i}$ is of Schmidt rank two. The largest subspace of
$\mathbbm{C}^3\ot\mathbbm{C}^3$ containing only vectors of
Schmidt-rank three is one-dimensional (see, e.g., Ref.
\cite{Montanaro}). On the other hand, if one of $\ket{\Psi_i}$
$(i=1,2)$ is of rank one, i.e., is product with respect to the
partition $AB|CD$,
$\bra{e,f^{*}}\rho^{T_{AB}}\ket{e,f^{*}}=\bra{e^*,f^*}\rho\ket{e^*,f^*}=0$
meaning that $\rho\ket{e^*,f^*}=0$. It is then clear that since
$\ket{e},\ket{f}\in \mathcal{S}_2$, $\mathcal{P}_4\ket{e^*,f^*}\in
K(\rho)$, implying that $r(\rho)=4$, which contradicts the
assumption.

Assuming then that $\ket{\Psi_1}$ is of rank two, either
$V_{AB,CD}\ket{\Psi_1^{*}}$ is linearly independent of
$\ket{\Psi_1}$ leading to Eq. (\ref{wektory}), or
$V_{AB,CD}\ket{\Psi_1^*}=\xi\ket{\Psi_1}$ for some
$\xi\in\mathbbm{C}$. In the latter case, short algebra implies
that $\ket{\Psi_i}$ $(i=1,2)$ are not linearly independent
contradicting the fact that they span two-dimensional kernel of
$\rho^{T_{AB}}$.

As a result, there is a vector
$\ket{e}=(1,\alpha)\in\mathbbm{C}^2$ such that $\ket{e^*}^{\ot
2}\ket{e}^{\ot 2}\in R(\rho^{T_{AB}})$ iff there is
$\alpha\in\mathbbm{C}$ solving the equation
\begin{equation}\label{equation0}
P(\alpha^{*})Q(\alpha)+\widetilde{P}(\alpha^{*})\widetilde{Q}(\alpha)=0,
\end{equation}
where $P,\widetilde{P}$ and $Q,\widetilde{Q}$ are polynomials
generically of degree two. Such $\alpha$ exists if and only if
there is $z\in\mathbbm{C}$ fulfilling
\begin{equation}\label{V1}
P(\alpha^{*})=z\widetilde{P}(\alpha^{*})
\end{equation}
and
\begin{equation}\label{V2}
\widetilde{Q}(\alpha)=-zQ(\alpha).
\end{equation}
With the aid of the first equation, we can determine $\alpha^*$ as
a function of $z$. There are clearly at most two such solutions.
Putting them to Eq. (\ref{V2}) and getting rid of the square root,
we arrive at a single equation
\begin{equation}\label{equation}
(z^{*})^{2}Q_{4}(z)+z^{*}Q'_4(z)+Q''_4(z)=0,
\end{equation}
where $Q_4$, $Q_4'$, and $Q_4''$ stand for polynomials which are
generically of fourth degree. It has at least one solution because
the polynomial appearing on the left-hand side of Eq.
(\ref{equation}) has unequal degrees in $z$ and $z^{*}$. This
allows for application of lemma \ref{lem_pol}, completing the
proof.
\end{proof}

One notices that this method cannot be directly applied in the
case of larger even $N$. Already for $N=6$, the left-hand side of
Eq. (\ref{equation0}) contains three terms and therefore the
factorization (\ref{V1}) and (\ref{V2}) cannot be done. Let us notice,
however, that the numerical search for extremal states of even $N$ done below
does not reveal examples of extremal PPT entangled symmetric states
of these ranks, suggesting the lack of edge states in generic states of these ranks.

More generally, it should be noticed that the analysis of edge states
allows for further reduction of configurations of ranks relevant
for characterization of PPT entanglement in symmetric states. This is because
a PPT state that is not edge can be written as a mixture of a pure product vector and
another PPT state of lower ranks (see also Sec. \ref{sec:special}).

\subsection{On the Schmidt number of symmetric states}

Let us finally comment on the Schmidt number of the symmetric
states. Clearly, a pure state $\ket{\psi}\in\mathcal{H}_N$ can be
written as a linear combination of fully product vectors from
$\mathcal{H}_N$. Following Ref. \cite{Schmidtnumber}, the smallest
number of terms in such decompositions of $\ket{\psi}$ is called
the Schmidt rank of $\ket{\psi}$ and denoted $r(\ket{\psi})$.
Then, analogously to Ref. \cite{TerhalHorodecki}, we can define
the Schmidt number of $\rho$ to be
$\min_{\{\ket{\psi_i}\}}\{\max_i r(\ket{\psi_i})\}$, where the
minimum is taken over all decompositions $\{\ket{\psi_i}\}$ of
$\rho$, i.e., $\rho=\sum_i\proj{\psi_i}$.

Below we show that in small symmetric systems consisting of four
or five qubits, any entangled state has the Schmidt number two or
at most three, respectively. We also comment on the Schmidt number
of larger systems.

Before that we need some preparation. Let us introduce the
following transformations: $F_n:(\mathbbm{C}^2)^{\ot n}\to
\mathbbm{C}^2$ and $G_n:(\mathbbm{C}^2)^{\ot (n+1)}\mapsto
(\mathbbm{C}^{2})^{\ot (2n+1)}$ defined through
\begin{equation}\label{Vn}
F_n(1,\alpha)^{\ot n}=(1,\alpha^n)
\end{equation}
and
\begin{equation}\label{Wn}
G_n[(1,\alpha^{n+1})\ot (1,\alpha)^{\ot n}]=(1,\alpha)^{\ot
(2n+1)},
\end{equation}
respectively, for any $\alpha\in\mathbbm{C}$ and $n=1,2,\ldots$.
Notice that both maps are of full rank. Then, by $\hat{F}_n$ and
$\hat{G}_n$ we denote maps that are defined through the adjoint
actions of $F_n$ and $G_n$, i.e., $\hat{X}(\cdot) =
X(\cdot)X^{\dagger}$ $(X = F_n,G_n)$.

Let us comment briefly on the properties of $F_n$ and $G_n$.
First, consider an $N$-qubit symmetric vector $\ket{\psi}$. An
application of $F_{\lceil N/2\rceil}$ to chosen $\lceil N/2\rceil$
qubits of $\ket{\psi}$, say the first ones,
brings it to an $(\lfloor N/2\rfloor+1)$-qubit vector
$\ket{\psi'}\in\mathbbm{C}^2\ot \mathcal{S}_{\lfloor N/2\rfloor}$.
A subsequent application of $G_{N/2-1}$ to the first $N/2$ qubits
of $\ket{\psi'}$ in case of even $N$ and $G_{\lceil N/2\rceil-1}$
to the whole $\ket{\psi'}$ in the case of odd $N$, returns
$\ket{\psi}$.

Analogously, by applying $\hat{F}_{\lceil N/2\rceil}$ to the first
$\lceil N/2\rceil$ qubits of an $N$-qubit mixed symmetric state
$\rho$, one brings it to an $(\lfloor N/2\rfloor+1)$-qubit state
$\sigma$ whose the last $\lfloor N/2\rfloor$ qubits are still
supported on the symmetric subspace $\mathcal{S}_{\lfloor
N/2\rfloor}$. Let us denote the parties of this state by
$B_1,\ldots,B_{\lfloor N/2\rfloor+1}$. With respect to the
bipartition $B_1|B_2\ldots B_{\lfloor N/2\rfloor+1}$, it can be
seen as a bipartite state acting on $\mathbbm{C}^2\ot
\mathbbm{C}^{\lfloor N/2\rfloor+1}$. Moreover, this ``compressing''
operation preserves the rank of any symmetric state, i.e.,
$r(\sigma)=r(\rho)$. This is because $R(\rho)$ is spanned by the
symmetric product vectors $(1,\alpha)^{\ot N}$ for
$\alpha\in\mathbbm{C}$, which are then mapped by $F_{\lceil
N/2\rceil}$ to $(1,\alpha)^{\ot \lfloor N/2\rfloor}\ot
(1,\alpha^{\lceil N/2\rceil})$. Since all powers of $\alpha$ from
the zeroth one to $\alpha^{N}$ still appear in the projected
vectors, the whole information about $\rho$ is encoded in
$\sigma$. Precisely, by an application of $\hat{G}_{N/2-1}$ to the
first $ N/2$ qubits of $\sigma$ for even $N$ and $\hat{G}_{\lceil
N/2\rceil-1}$ to all qubits for odd $N$ of $\sigma$, one recovers
$\rho$. With all this we are now ready to state and prove the
following theorem.

\begin{thm}\label{thm_decom}
Let $\rho$ be an entangled $N$-qubit symmetric state. If by an
application of $\hat{F}_{\lceil N/2\rceil}$ to the first $\lceil
N/2\rceil$ qubits of $\rho$ one gets an $(\lfloor N/2\rfloor+1)$-qubit
state $\sigma$ (see above) that is separable with
respect to the bipartition $B_1|B_2\ldots
B_{\lfloor N/2\rfloor+1}$,
then $\rho$ can be written as
\begin{equation}\label{wielka_dupa0}
\rho=\sum_{i=1}^{K}\left[\sum_{j=1}^{\lceil
N/2\rceil}A_j^{(i)}(1,\alpha_{j}^{(i)})^{\ot N}\right],
\end{equation}
where $A_j^{i},\alpha_j^{i}\in\mathbbm{C}$ and $[\psi]$ denotes a
projector onto $\ket{\psi}$.
\end{thm}
\begin{proof}We can clearly assume that $r(\rho)=N+1$. Let the
$(\lfloor N/2\rfloor+1)$-qubit state $\sigma$, coming from the
application of $\hat{F}_{\lceil N/2\rceil}$ to the first $\lceil
N/2\rceil$ qubits of $\rho$, be separable with respect to the
Hilbert space $\mathbbm{C}^2\ot\mathbbm{C}^{\lfloor
N/2\rfloor+1}$. It can then be written as
\begin{equation}\label{dupa}
\sigma=\sum_{i=1}^{K}p_i\proj{e_i}\ot\proj{f_i},\qquad
p_i\geq0,\qquad \sum_ip_i=1,
\end{equation}
where $\ket{e_i}\in\mathbbm{C}^{2}$ and $\ket{f_i}\in
\mathbbm{C}^{\lfloor N/2\rfloor+1}\cong \mathcal{S}_{\lfloor
N/2\rfloor}$.

As already said, the mapping $\hat{F}_{\lceil N/2\rceil}$
preserves the rank, and therefore $r(\sigma)=r(\rho)=N+1$.
On the other hand, $\sigma$ acts on
$\mathbbm{C}^2\ot\mathbbm{C}^{\lfloor N/2\rfloor+1}$ and therefore
its maximal rank is $2(\lfloor N/2\rfloor+1)$, which means that
for odd $N$, $\sigma$ is of full rank, while for even $N$ it has a
single vector $\ket{\phi}=\ket{10}-\ket{0,N/2}$ in its kernel.

Let us now divide our considerations into the cases of odd and
even $N$. In the first case we put $\ket{e_i}=(1,\alpha_i^{\lceil
N/2\rceil})$, and then notice that the isomorphism
$\mathbbm{C}^{\lfloor N/2\rfloor+1}\cong \mathcal{S}_{\lfloor
N/2\rfloor}$ implies that each $\ket{f_i}\in \mathbbm{C}^{\lfloor
N/2\rfloor+1}$ can be expanded in terms of $\lfloor N/2\rfloor+1$
product symmetric vectors from $\mathcal{S}_{\lfloor N/2\rfloor}$
in the following way
\begin{equation}
\ket{f_i}=\sum_{j=1}^{\lfloor
N/2\rfloor+1}A_j^{(i)}(1,\mathrm{e}^{\mathrm{i}\varphi_j}\alpha_i)^{\ot
\lfloor N/2\rfloor}
\end{equation}
with
\begin{equation}
\varphi_j=\frac{2\pi j}{\lceil N/2\rceil},
\end{equation}
and $A_j^{(i)}\in\mathbbm{C}$. Consequently,
\begin{equation}
G_{{\lceil N/2\rceil}-1}\ket{e_i}\ket{f_i}=\sum_{j=1}^{\lfloor
N/2\rfloor+1}A_j^{(i)}(1,\mathrm{e}^{\mathrm{i}\varphi_j}\alpha_i)^{\ot
N},
\end{equation}
which, when substituted to Eq. (\ref{dupa}) leads directly to
(\ref{wielka_dupa0}).

In the case of even $N$, $\sigma$ has a single vector
$\ket{\phi}=\ket{01}-\ket{N/2,0}$ in $K(\sigma)$. Putting again
$\ket{e_i}=(1,\alpha_i^{N/2})$ with $\alpha_i\in\mathbbm{C}$, it
imposes a constraint on $\ket{f_i}$, that is, $\langle
N/2|f_i\rangle=\langle 0|f_i\rangle \alpha^{N/2}_i$, meaning that
$\ket{f_i}\in\mathbbm{C}^{N/2}$. Therefore, again exploiting the
isomorphism $\mathbbm{C}^{N/2}\cong \mathcal{S}_{N/2-1}$, one sees that all
$\ket{f_i}$ can be written as
\begin{equation}
\ket{f_i}=\sum_{j=1}^{
N/2}A_j^{(i)}(1,\mathrm{e}^{\mathrm{i}\varphi_j}\alpha_i)^{\ot
N/2}.
\end{equation}
By substituting this to Eq. (\ref{dupa}) and applying
$\hat{G}_{N/2-1}$ to $\sigma$, one gets Eq. (\ref{wielka_dupa0}),
completing the proof.

Let us finally comment on the choice of the product symmetric
vectors used to expand $\ket{f_i}$'s. In both cases of odd and
even $N$ they are chosen to be
$(1,\mathrm{e}^{\mathrm{i}\varphi_j}\alpha)^{\ot \lfloor
N/2\rfloor}$ $(j=0,\ldots,\lceil N/2\rceil-1)$, where $\varphi_j$
are chosen so that $\mathrm{e}^{\mathrm{i}\varphi_j}$ are $\lceil
N/2\rceil$th roots of the unity. One checks by hand that such
vectors span a $\lceil N/2\rceil$-dimensional linear space.
\end{proof}

Theorem \ref{thm_decom} implies that any PPT entangled symmetric state
consisting of four (five) qubits has Schmidt number two (at most
three). To be more precise, we prove the following corollaries.

\begin{cor}\label{cor:45}
Any PPT entangled symmetric state $\rho$ of four qubits can be
written as
\begin{equation}\label{wielka_dupa}
\rho=\sum_{i=1}^{K} [A^{(i)}_1 (1,\alpha_i)^{\ot 4}+A^{(i)}_2
(1,-\alpha_i)^{\ot 4}],
\end{equation}
while any PPT entangled symmetric state of five qubits as
\begin{eqnarray}\label{wielka_dupa2}
\rho&=&\sum_{i=1}^{K} [A^{(i)}_1 (1,\alpha_i)^{\ot 5}+A^{(i)}_2
(1,\mathrm{e}^{\mathrm{i}\frac{2\pi}{3}}\alpha_i)^{\ot
5}\nonumber\\
&&\qquad+A^{(i)}_3(1,\mathrm{e}^{\mathrm{i}\frac{4\pi}{3}}\alpha_i)^{\ot
5}],
\end{eqnarray}
with $K\leq 6$ and $A_j^{(i)},\alpha_i\in\mathbbm{C}$.
\end{cor}
\begin{proof}
By applying $\hat{F}_2$ ($\hat{F}_3$) to the first two (three)
qubits of $\rho$ for $N=4$ $(N=5)$ we get a state $\sigma$ acting
on $\mathbbm{C}^2\ot\mathbbm{C}^3$. It is clearly PPT and due to
Ref. \cite{2N} also separable. Consequently, theorem
\ref{thm_decom} implies that $\rho$ can be written as in Eq.
(\ref{wielka_dupa0}), which in particular cases of $N=4$ and $N=5$
leads to (\ref{wielka_dupa}) and (\ref{wielka_dupa2}),
respectively. The number of elements in both the decompositions
(\ref{wielka_dupa}) and (\ref{wielka_dupa2}) follows from the fact
that any qubit-qutrit separable state can be written as a convex
combination of six product vectors \cite{Normal}.
\end{proof}

It should be noticed that by using the approach developed in Ref.
\cite{Normal}, one can obtain decompositions of any PPT entangled
four-qubit symmetric state similar to (\ref{wielka_dupa}), but in
which vectors $(1,-\alpha_i)^{\ot 4}$ are replaced by either
$(0,1)^{\ot 4}$ or $(1,0)^{\ot 4}$.

\begin{thm}\label{thm_dec22}
Let $\rho$ be a PPT entangled symmetric four-qubit state. Then it
can be written as
\begin{equation}\label{Lemma13}
\rho=\sum_{i=1}^{K} [A_i (1,\alpha_i)^{\ot 4}+B_i (0,1)^{\ot 4}],
\end{equation}
where $K\leq 6$, $(1,\alpha_i)\in\mathbbm{C}^{2}$, and $A_i,B_i$
are some complex coefficients, and by $[\psi]$ we denote a
projector onto $\ket{\psi}$.
\end{thm}
\begin{proof}
The proof exploits the method developed in Ref. \cite{Normal}.
First, one notices that any $\rho$ can be written as a sum of
rank-one matrices
\begin{equation}\label{ensemble}
\rho=\sum_{i=1}^{K}\proj{\Psi_i},
\end{equation}
where, in particular, $\ket{\Psi_i}$ can be (unnormalized)
eigenvectors of $\rho$, and $K\leq 6$ (see corollary
\ref{cor:45}). On the other hand, $\rho$ can always be expressed
in terms of the symmetric unnormalized basis
$\{\ket{E^4_{\mu}}\}_{\mu=1}^{5}$ spanning $\mathcal{S}_4$ as
\begin{equation}\label{decom2}
\rho=\sum_{\mu,\nu=1}^{5}\rho_{\mu\nu}\ket{E^4_{\mu}}\!\bra{E^4_{\nu}}.
\end{equation}
Both decompositions (\ref{ensemble}) and (\ref{decom2}) are
related {\it via} the so-called Gram system of $\rho$, i.e., a
collection of $K$-dimensional vectors $\ket{v_{\mu}} = (1/\langle
E^4_{\mu}|E^4_{\mu}\rangle)(\langle
\Psi_1|E^4_{\mu}\rangle,\ldots,\langle \Psi_K|E^4_{\mu}\rangle)$
$(\mu=1,\ldots,5)$, giving $\rho_{\mu\nu}=\langle
v_{\mu}|v_{\nu}\rangle$. Putting the latter to Eq. (\ref{decom2})
with explicit forms of the vectors $\ket{v_{\mu}}$, one recovers
(\ref{ensemble}).

Now, by projecting the last party onto $\ket{0}$ we get a
three-qubit symmetric PPT state $\widetilde{\rho}$, which, as
already stated,  is separable. Then, according to Ref.
\cite{Normal}, there exists a diagonal matrix
$M=\mathrm{diag}[\alpha_1^{*},\ldots,\alpha_K^{*}]$ such that
$\ket{v_{\mu}}=M^{\mu-1}\ket{v_{1}}$ $(\mu=1,\ldots,4)$. For
convenience we can also put
$\ket{v_{5}}=M^{4}\ket{v_1}+\ket{\widetilde{v}}$ with
$\ket{\widetilde{v}}$ being some $K$-dimensional complex vector.
Then, putting $\ket{v_1}=(A_1^{*},\ldots,A_K^{*})$ and
$\ket{\widetilde{v}}=(B_1^{*},\ldots,B_K^{*})$, one sees that
\begin{eqnarray}
\ket{\Psi_i}&=&\sum_{\mu=1}^{5}\frac{\langle E^4_{\mu}|\Psi_i\rangle}{\langle E^4_{\mu}|E^4_{\mu}\rangle}\ket{E^4_{\mu}}\nonumber\\
&=&A_i\sum_{\mu=1}^{4}\alpha_{i}^{\mu-1}\ket{E^4_{\mu}}+B_i\ket{E^4_5}\nonumber\\
&=&A_i(1,\alpha_i)^{\ot 4}+B_i\ket{E^4_5},
\end{eqnarray}
where the second equation follows from the explicit form of the
vectors $\ket{v_{\mu}}$. Substituting vectors $\ket{\Psi_i}$ to
Eq. (\ref{ensemble}), one gets (\ref{Lemma13}), which completes
the proof.
\end{proof}

In order to get the representation (\ref{Lemma13}) with $(0,1)$
replaced by $(1,0)$, one has to project the last party of $\rho$
onto $\ket{1}$ instead of $\ket{0}$.

\section{Extremal PPT entangled symmetric states}
\label{sec:Extr}

We will here seek extremal elements in the convex set of $N$-qubit
symmetric PPT states $D_{\mathrm{PPT}}^{\mathrm{sym}}$. We see
that the number of distinct configurations of ranks for which one
finds such examples is small and does not increase with $N$. In
particular, if $N$ is even there is only a single such
configuration.

Let us start by adapting to the multipartite case an algorithm
described in Ref. \cite{Leinaas} (see also. Ref. \cite{OptComm})
allowing to look for extremal elements of $D_{\mathrm{PPT}}$.

\subsection{An algorithm allowing to seek multipartite extremal PPT states}
\label{sec:algor}

Let us consider again a product Hilbert space
$\mathcal{H}_N=\mathbbm{C}^{d_1}\ot\ldots\ot\mathbbm{C}^{d_1}$ and
the set $D_{\mathrm{PPT}}$ of fully PPT states acting on
$\mathcal{H}_N$. We call an element $\rho$ of $D_{\mathrm{PPT}}$
{\it extremal} iff it does not allow for the decomposition
\begin{equation}\label{extr}
\rho=p\rho_1+(1-p)\rho_2
\end{equation}
with $\rho_i\in D_{\mathrm{PPT}}$ such that $\rho_1\neq\rho_2$ and
$0<p<1$. Should $\rho$ be not extremal, Eq. (\ref{extr}) implies
that $R(\rho_1)\subseteq R(\rho)$ and $R(\rho_2)\subseteq R(\rho)$
and also $R(\rho_i^{T_k})\subseteq R(\rho^{T_k})$ for all $k$.
Alternatively speaking, if $\rho$ is not extremal one is able to
find another PPT state $\sigma\neq \rho$ such that
\begin{equation}\label{ran}
R(\sigma)\subseteq R(\rho)
\end{equation}
and
\begin{equation}\label{ranTk}
R(\sigma^{T_k})\subseteq R(\rho^{T_k}) \qquad (k=1,\ldots,M),
\end{equation}
where by $M$ we denote the smallest number of partial transpositions necessary to
define the set of PPT states.

On the other hand, if there is a PPT density matrix $\sigma\neq
\rho$ fulfilling the conditions (\ref{ran}) and (\ref{ranTk}),
then one finds such a number $x>0$ that
$\rho(x)=(1+x)\rho-x\sigma$ is again a PPT state. Thus, $\rho$
cannot be extremal because it admits the decomposition
(\ref{extr}), i.e., $\rho=[\rho(x)+x\sigma]/(1+x)$. In this way,
one arrives at a natural criterion for extremality saying that
$\rho$ is extremal iff there is no state $\sigma\neq \rho$ such that
(\ref{ran}) and (\ref{ranTk}) hold.

Interestingly, one can even relax the assumption of $\sigma$ being
a PPT state to being a Hermitian matrix satisfying a set of equations
following from Eqs. (\ref{ran}) and (\ref{ranTk}):
\begin{equation}\label{cond_extr}
\mathcal{P}_k h^{T_k}\mathcal{P}_k=h^{T_k}\qquad (k=0,\ldots,M),
\end{equation}
where $T_0$ is an identity map. This is because if such $h$ exists,
one considers a one-parameter class of matrices $\rho(x)=(1+x\Tr h)\rho-xh$
($\Tr h$ is added in order to ensure that $\Tr[\rho(x)]=1$ for any $x$). It is
straightforward to show that there exist $x_1<0$ and $x_2>0$ such
that for any $x\in[x_1,x_2]$, $\rho(x_i)$ $(i=1,2)$ are PPT
states. Consequently, $\rho=[1/(|x_1|+x_2)][x_2\rho(x_1)+x_1\rho(x_2)]$,
meaning that it cannot be extremal.

Let us also notice that the system (\ref{cond_extr}) is equivalent to a
single equation
\begin{equation}\label{cond_extr2}
[\hat{\mathcal{P}}_M\circ\ldots\circ\hat{\mathcal{P}}_1\circ\hat{\mathcal{P}} ](h)=h,
\end{equation}
where the maps $\hat{\mathcal{P}}_k$ are defined through
$\hat{\mathcal{P}}_k(\cdot)=[\mathcal{P}_k(\cdot)^{T_k}\mathcal{P}_k]^{T_k}$ $(k=0,\ldots,M)$.
Indeed, if $h$ satisfies the system (\ref{cond_extr}) then it also
obeys (\ref{cond_extr2}). On the other hand, if the condition
(\ref{cond_extr2}) is satisfied with some $h$, then it has to obey
(\ref{cond_extr}); if one of the conditions (\ref{cond_extr}) does
not hold, a simple comparison of norms of both sides of
(\ref{cond_extr2}) shows that (\ref{cond_extr}) neither can be
satisfied.

As a result, we have just reached an operational criterion for
extremality of elements of $D_{\mathrm{PPT}}$ \cite{Leinaas} (see
also Ref. \cite{OptComm}).
\begin{thm}\label{thm_extr}
A given PPT state $\rho$ acting on $\mathcal{H}_N$ is extremal in
$D_{\mathrm{PPT}}$ iff there does not exists a Hermitian solution to the system
(\ref{cond_extr}) which is linearly independent of $\rho$.
\end{thm}

This also leads to a necessary condition for extremality that can
be formulated in terms of the ranks $r(\rho^{T_k})$
$(k=1,\ldots,M)$. Namely, each equation in (\ref{cond_extr})
imposes $[\dim \mathcal{H}_N]^2-[r(\rho^{T_k})]^2$ linear
constraints on the matrix $h$. The maximal number of the
constraints imposed by the system is thus $\sum_{k=0}^M([\dim
\mathcal{H}_{N}]^2-[r(\rho^{T_k})]^2)$. On the other hand, a
Hermitian matrix acting on $\mathcal{H}_{N}$ is specified by
$[\dim \mathcal{H}_{N}]^2$ real parameters and therefore if
\begin{equation}\label{nec_cond}
\sum_{k}[r(\rho^{T_k})]^2\geq M[\dim\mathcal{H}_N]^2+1,
\end{equation}
the system (\ref{cond_extr}) has a solution and $\rho$ is not
extremal.

Importantly, the above considerations imply an algorithm allowing
to seek extremal elements of $D_{\mathrm{PPT}}$ \cite{Leinaas}.
Given a PPT state $\rho$ and a solution $h$ to the system (\ref{cond_extr}), one
considers $\rho(x)=(1+x\Tr h)\rho-xh$. It is fairly easy to see
that there is $x=x_*$ for which $\rho_1\equiv\rho(x_*)\in
D_{\mathrm{PPT}}$ but $r(\rho_1^{T_k})=r(\rho^{T_k})-1$ for some
$k$. In other words, we can choose $x$ in such way that one of the
ranks of the resulting state $\rho_1$ is diminished by one.

For the resulting state $\rho_1$ we again look for solutions to
(\ref{cond_extr}). If the only solution is $\rho_1$ itself (up to normalization),
then it is already extremal. If not, one again considers
$\rho_1(x)=(1+x\Tr h)\rho_1-xh$ and finds such $x=x_*$ that
$\rho_2\equiv \rho_1(x_*)$ is a PPT state with one of the ranks
diminished by one. We proceed in this way until we obtain an
extremal state. If the latter has rank one it is separable,
otherwise it has to be entangled. Clearly, since we deal with
finite-dimensional Hilbert space, the final state of this
algorithm is reached in a finite number of steps.

In our implementation of this algorithm we obtain the Hermitian
matrix $h$ by solving a slightly different system of equations
than (\ref{cond_extr}) (or the single one (\ref{cond_extr2})), i.e.,
\begin{equation}\label{implementation}
h^{T_k}\ket{\Psi_i^{(k)}}=0\qquad (i=1,\ldots,k(\rho^{T_k}))
\end{equation}
for all $k=0,\ldots,M$ with $\ket{\Psi_i^{(k)}}$ denoting vectors
spanning the kernel of $\rho^{T_k}$. Clearly, both systems
(\ref{cond_extr}) and (\ref{implementation}) are equivalent.

A general Hermitian matrix acting on $\mathcal{H}_N$ is fully
characterized by $(\dim\mathcal{H}_N)^2$ real parameters $h_i$,
which we consider elements of a vector $\ket{h}\in\mathbbm{C}^{(\dim\mathcal{H}_N)^2}$.
Then, the set
(\ref{implementation}) can be easily reformulated as a single
matrix equation $\mathcal{R}\ket{h}=0$, where $\mathcal{R}$ is a
$\big[2\dim\mathcal{H}_N\sum_{l=0}^{M}k(\rho^{T_l})\big]\times (\dim\mathcal{H}_N)^2$
matrix with real entries. The number of rows of $\mathcal{R}$
stems from the fact that each group of equations in
(\ref{implementation}) corresponding to a given $k$ gives $2
k(\rho^{T_k})\dim\mathcal{H}_N$ linear and real conditions for $h_i$, which
when summed over all partial transpositions (including also $T_0$)
results in the aforementioned number of rows. Clearly, this gives more conditions for $h_i$ than
those following from Eqs. (\ref{cond_extr}), i.e.,
$\sum_{k=0}^{M}[(\dim\mathcal{H})^2-[r(\rho^T_k)]^2]$ (some of them are clearly linearly dependent).
Nevertheless, even if the number of equations is larger in comparison to Eq.
(\ref{cond_extr2}), our approach does not require multiplying many
matrices, which in the case of many parties makes the
implementation a bit faster.

\subsection{Extremal PPT entangled symmetric states}
\label{sec:SymmExtr}

Let us now apply the above considerations to the symmetric states.
First of all, one notices that the condition (\ref{nec_cond}) has
to be slightly modified because different partial transpositions of
a symmetric state and the state itself act on Hilbert spaces of
different dimensions (see Ref. \cite{OptComm} for the case of four-qubits).
Specifically, an equation in
(\ref{cond_extr}) corresponding to the partial transposition with
respect to $S$ gives $(|S|+1)^2(N-|S|+1)^2-[r(\rho^{T_S})]^2$
$(|S|=0,\ldots,\lfloor N/2\rfloor)$ linear equations. Then, the
condition (\ref{nec_cond}) becomes
\begin{equation}
\sum_{|S|=0}^{\lfloor N/2\rfloor}[r(\rho^{T_S})]^2\geq
\sum_{|S|=0}^{\lfloor N/2\rfloor}(|S|+1)^2(N-|S|+1)^2-(N+1)^2+1.
\end{equation}
Taking into account the fact that among symmetric states only those of rank $N+1$ can be
entangled, allows us to rewrite the above inequality as
\begin{equation}\label{Extr_Ineq2}
\sum_{|S|=1}^{\lfloor N/2\rfloor}[r(\rho^{T_S})]^2\geq
\sum_{|S|=1}^{\lfloor N/2\rfloor}(|S|+1)^2(N-|S|+1)^2-(N+1)^2+1.
\end{equation}

In table \ref{Excl_Ranks} we collect all the ranks, singled out
with the aid of Ineq. (\ref{Extr_Ineq2}), for which there are no
extremal PPT states for exemplary cases of low number of qubits
$N=4,5,6$.

\begin{table}[h!]
\begin{tabular}{c|l|l}
  \hline
$N$    & Inequality (\ref{Extr_Ineq2}) & Ranks excluded with (\ref{Extr_Ineq2})\\
  \hline\hline

4 & $[r(\rho^{T_1})]^2+[r(\rho^{T_2})]^2\geq 121$ &
$(5,7,9),(5,8,8),(5,8,9)$ 

\\ \hline

5 & $[r(\rho^{T_1})]^2+[r(\rho^{T_2})]^2\geq 209$ &
$(6,9,12),(6,10,11),(6,10,12)$ 
\\ \hline

6 & $\begin{array}{l}
[r(\rho^{T_1})]^2+[r(\rho^{T_2})]^2\\+[r(\rho^{T_3})]^2\geq
577\end{array}$ & $\begin{array}{l} \mathrm{(7,10,15,16),(7,11,15,16),}\\

\mathrm{(7,12,14,16),
(7,12,15,15),}\\\mathrm{(7,12,15,16)}\end{array}$
\\ \hline\hline

\end{tabular}
\caption{Inequality (\ref{Extr_Ineq2}) (second column) for
exemplary cases of $N=4,5,6$ together with the ranks (third
column) excluded with its aid for which there are no extremal
entangled symmetric states. Notice that the fact that there are no
extremal states of maximal ranks can be inferred without restoring
to Ineq. (\ref{Extr_Ineq2}). }\label{Excl_Ranks}
\end{table}

We have applied the above algorithm to the PPT symmetric states
with the number of qubits varying from $N=4$ to $N=23$. As the
initial state we took the projector onto the symmetric space
$\mathcal{P}_N$. Clearly, it has all the ranks maximal, which is
important from the point of view of the algorithm; if the ranks of
the initial state $\rho$ are too low (cf. theorems \ref{thm_sep2}
and \ref{sep_gen}), the algorithm cannot produce an entangled PPT
state out of $\rho$ by lowering its ranks. Then, we have searched
for random extremal states by choosing randomly at each stage of
the protocol the matrices $h$ resulting from solving
(\ref{cond_extr}). On the other hand, to find other examples of
ranks than those obtained through a random search and not excluded
by the analysis above, we designed the matrices $h$ in such way
that they lower specific ranks. The obtained ranks are collected
in table \ref{Extr_Ranks}. Interestingly, there are always at most
three different configurations of ranks assumed by the found
extremal PPT entangled symmetric states and it seems that the
number of configurations does not increase with $N$. Moreover, in
the case of odd $N$, there is only a single such configuration
(all ranks are maximal except for the last one which is two less
than maximal). It is an interesting problem to confirm these
findings analytically. And if this is the case, the problem of
characterization of PPT entanglement in symmetric states reduces
significantly to the characterization of extremal states assuming
few different configurations of ranks, in particular a single one
for odd $N$. Notice that in the case of symmetric qubits, there
cannot be PPT entangled states of lower ranks than those assumed
by extremal one as this is the case in higher-dimensional Hilbert
spaces (cf. Ref. \cite{Leinaas2}). This is because in order to
construct such states one needs PPT extremal entangled states
supported on lower-dimensional Hilbert spaces, and in our case
such states are always separable.

Let us study in detail extremal entangled states in the exemplary case of $N=4$.
From theorems \ref{thm_sep2} and \ref{sep_gen} it follows that PPT states of ranks
$(5,r(\rho^{T_A}),r(\rho^{T_{AB}}))$ with $r(\rho^{T_A})\leq 6$ or $r(\rho^{T_{AB}})\leq 6$
are either all separable or generically separable. Then, theorem \ref{thm:edge2}
states that generic PPT states of ranks $(5,8,7)$ are not edge and thus not extremal.
Finally, Ineq. (\ref{cond_extr2}) implies that PPT states of ranks $(5,7,9)$,
$(5,8,8)$ and $(5,8,9)$ cannot be extremal. As a results, the natural candidates for
extremal states that can be obtained with the aid of the above algorithm
have ranks $(5,7,7)$ and $(5,7,8)$.

We have run the algorithm $30000$ times and $19.2\%$ of the generated
examples were extremal entangled states of ranks $(5,7,8)$.
In the remaining $80.8\%$ of cases we arrived at states of ranks
$(5,7,7)$, all being separable. Also, when lowering the ranks from the
initial state of ranks $(5,8,9)$, $99.4\%$
of the times we have obtained an intermediate $(5,8,8)$-state,
whereas intermediate states of ranks $(5,7,9)$
have appeared the $0.6\%$ of remaining cases.
Concluding, it should be noticed that with the aid of the
above algorithm we have generated PPT entangled extremal
states assuming only a single configuration of ranks.
All states of ranks $(5,7,7)$ occurred to be separable and there is an indication
suggesting that generic four-qubit PPT symmetric states of these ranks are separable
(see appendix \ref{app}).

Let us finally notice that to make the application of the algorithm to
systems consisting of even twenty-three qubits possible, one has to take
an advantage of the underlying symmetry and try to avoid
representing a symmetric $\rho$ and its partial transpositions in the
full Hilbert space $\mathcal{H}_{2,N}=(\mathbbm{C}^2)^{\ot N}$
(see also Ref. \cite{Stockton}). Indeed, since
$\mathcal{S}_N\cong\mathbbm{C}^{N+1}$, one can represent $\rho$
or a general Hermitian matrix supported on $\mathcal{S}_N$, as a
$(N+1)\times(N+1)$ matrix, which we further denote
$\rho_{\mathrm{red}}$. In order to move from one representation to the
other one we use a $(N+1)\times 2^N$ matrix
$B_N:\mathcal{S}_N\mapsto\mathbbm{C}^{N+1}$ given by
\begin{equation}
B_N=\sum_{m=0}^{N}\ket{m}\!\bra{\widetilde{E}_{m}^{N}},
\end{equation}
which gives $\rho_{\mathrm{red}}=B_N\rho B_N$. It is
straightforward to check that for any $N$,
$B^{T}_{N}B_{N}=\mathcal{P}_N$ and $B_NB^{T}_N=\mathbbm{1}_{N+1}$.

Then, accordingly, the partial transposition of $\rho$ with
respect to $T_k$ can be represented as a
$(k+1)(N-k+1)\times(k+1)(N-k+1)$ matrix
$\rho_{\mathrm{red}}^{T_k}$ acting on
$\mathbbm{C}^{k+1}\ot\mathbbm{C}^{N-k+1}$. To get the latter from
$\rho_{\mathrm{red}}$ without restoring to the representation of
$\rho$ in the full Hilbert space $\mathcal{H}_{2,N}$, one can
utilize a $(k+1)(N-k+1)\times (N+1)$ matrix
$\widetilde{B}_k=(B_k\ot B_{N-k})B_{N}^{T}$, i.e.,
\begin{eqnarray}\label{kolejne}
\hspace{-0.7cm}\rho_{\mathrm{red}}^{T_k}&=&[(B_k\ot B_{N-k})\rho(B_k^T\ot B_{N-k}^T)]^{T_k}\nonumber\\
&=&[(B_k\ot B_{N-k})B_N^{T}\rho_{\mathrm{red}}B_N(B_k^T\ot
B_{N-k}^T)]^{T_k}\nonumber\\
&=&(\widetilde{B}_k\rho_{\mathrm{red}}\widetilde{B}_k^{T})^{T_k}.
\end{eqnarray}
Short algebra shows that the elements of $\widetilde{B}_k$
$(k=1,\ldots,\lfloor N/2\rfloor)$ are given by
\begin{equation}
\bra{i,j}\widetilde{B}_k\ket{n}=\sqrt{\binom{N}{i}\binom{N}{j}/\binom{N}{n}}\,\,\delta_{i+j=n}.
\end{equation}
with $i=0,\ldots,k$, $j=0,\ldots,N-k$, and $n=0,\ldots,N$.

Consequently, to get effectively a partial transposition of
$\rho$, one maps a $(N+1)\times (N+1)$ matrix
$\rho_{\mathrm{red}}$ with $\widetilde{B}_k$, and subsequently
performs a simple partial transposition on the resulting bipartite
matrix [cf. Eq. (\ref{kolejne})]. Accordingly, one also transforms
the Hermitian matrices $h$ appearing in the system
(\ref{implementation}). Notice that this approach allows us to
reduce the algorithm complexity from exponential to polynomial in
$N$ both in time and memory. Precisely, the estimated time
complexity of our approach amounts to $O(N^6)$.

\begin{widetext}

\begin{table}[h!]
\centering
\begin{tabular}{|c||c|c|c|c|c|c|c|c|c|c|c|c|}
%
$N$&$r(\rho)$&$r(\rho^{\Gamma_1})$&$r(\rho^{\Gamma_2})$&$r(\rho^{\Gamma_3})$&$r(\rho^{\Gamma_4})$&$r(\rho^{\Gamma_5})$
&$r(\rho^{\Gamma_6})$&$r(\rho^{\Gamma_7})$&$r(\rho^{\Gamma_8})$&$r(\rho^{\Gamma_9})$&$r(\rho^{\Gamma_{10}})$&
$r(\rho^{\Gamma_{11}})$\\
\hline \hline
4&  5&  7 \quad (-1) &  8\quad (-1)&&&&&&&&&\\
\hline
5&  6&  10&     10 \quad (-2) &&&&&&&&&\\
\hline
6&  7&  12& 14 \quad (-1) &     14 \quad (-2) &&&&&&&&\\
 &   &    &     14 \quad (-1) &     13 \quad (-3) &&&&&&&&\\
\hline
7 &     8&  14& 18& 18 \quad (-2)&&&&&&&&\\
\hline
8&9&16&21&  23 \quad (-1) & 23 \quad (-2)&&&&&&&\\
 & &  &  &  23 \quad (-1) & 22 \quad (-3)&&&&&&&\\
\hline
9&10&18&24&28&28 \quad (-2)&&&&&&&\\
\hline
10&11&20&27&32&34 \quad (-1)&33 \quad (-3)&&&&&&\\
  &  &  &  &  &34 \quad (-1)&34 \quad (-2)&&&&&&\\
  &  &  &  &  &35 \quad (+0)&32 \quad (-4)&&&&&&\\
\hline
11&12&22&30&36&40&40\quad(-2)&&&&&&\\
\hline
12&13&24&33&40&45&47 \quad (-1)&47\quad (-2)&&&&&\\
  &  &  &  &  &  &47 \quad (-1)&46\quad (-3)&&&&&\\
  &  &  &  &  &  &48 \quad (+0)&45\quad (-4)&&&&&\\
\hline
13&14&26&36&44&50&54&54\quad (-2)&&&&&\\
\hline

14&15&28&39&48&55&60&62 \quad (-1)&62 \quad (-2)&&&&\\
  &  &  &  &  &  &  &62 \quad (-1)&61 \quad (-3)&&&&\\
  &  &  &  &  &  &  &63\quad (+0)&60\quad(-4)&&&&\\
\hline
15&16&30&42&52&60&66&70&70 \quad (-2)&&&&\\
\hline
16&17&32&45&56&65&72&77&79\quad (-1)&79\quad (-2)&&&\\
  &  &  &  &  &  &  &  &79\quad (-1)&78\quad (-3)&&&\\
  &  &  &  &  &  &  &  &80\quad (+0)&77\quad (-4)&&&\\
\hline
17&18&34&48&60&70&78&84&88&88\quad (-2)&&&\\
\hline
18&19&36&51&64&75&84&91&96&98 \quad (-1)&98 \quad (-2)&&\\
  &  &  &  &  &  &  &  &  &98 \quad (-1)&97 \quad (-3)&&\\
  &  &  &  &  &  &  &  &  &99\quad (+0)&96\quad (-4)&&\\
\hline
19&20&38&54&68&80&90&98&104&108&108\quad (-2)&&\\
\hline

20&21&40&57&72&85&96&105&112&117&119\quad (-1)&119\quad (-2)&\\
  &  &  &  &  &  &  &   &   &   &119\quad (-1)&118\quad (-3)&\\
  &  &  &  &  &  &  &   &   &   &120\quad (+0)&117\quad (-4)&\\
\hline
21&22&42&60&76&90&102&112&120&126&130&130\quad (-2)&\\
\hline
22&23&44&63&80&95&108&119&128&135&140&142\quad (-1)&142\quad (-2)\\
  &  &  &  &  &  &   &   &   &   &   &142\quad (-1)&141\quad (-3)\\
  &  &  &  &  &  &   &   &   &   &   &143\quad (+0)&140\quad (-4)\\
\hline
23&24&46&66&84&100&114&126&136&144&150&154&154\quad (-2)\\
\hline
%
\end{tabular}
\caption{The ranks of extremal states found by using the algorithm
described in Sec. \ref{sec:algor}. The first column contains the
number of qubits, while the next six columns the ranks of $\rho$
and its partial transpositions $r(\rho^{T_i})$ $(i=1,\ldots,11)$.
Notice that there are no PPT entangled states with less qubits
than four \cite{EckertAnP} (cf. Sec. \ref{sec:sep}). The negative
numbers in parentheses denote the difference between the given
rank and its maximal value (the lack of parentheses means that the
rank is maximal). For all $N$ there are at most three possible
configurations of ranks with extremal states, and, interestingly,
in the case of odd $N$ there is always only one such
configuration. }\label{Extr_Ranks}
\end{table}
\end{widetext}

\section{Special cases}
\label{sec:special}
Here we summarize the obtained results for particular systems
consisting of four, five, and six qubits.

\textit{N=4}. It follows from theorem \ref{thm_sep2} that the four-qubit
symmetric states are separable if either $r(\rho)\leq 4$ or
$r(\rho^{T_A})\leq 4$, or $r(\rho^{T_{AB}})\leq 3$. Then, theorem
\ref{sep_gen} implies that if either $r(\rho^{T_A})\leq 6$ or
$r(\rho^{T_{AB}})\leq 6$, generic symmetric $\rho$ is separable.
This leaves only six configuration of ranks (out of all possible
72 assuming that $r(\rho)=5$) among which one may seek PPT entangled symmetric states:
$(5,7,7)$, $(5,7,8)$, $(5,7,9)$, $(5,8,7)$, $(5,8,8)$, and
$(5,8,9)$.

Passing to edgeness, it is known that states of ranks $(5,8,9)$
cannot be edge. Then, it follows from theorem \ref{thm:edge1} that
all states of ranks $(5,8,8)$ are not edge, while from theorems
\ref{thm:edge0} and \ref{thm:edge2} that generic states of
ranks $(5,7,9)$ and $(5,8,7)$ are not edge. These
theorems also show that a typical PPT entangled state
assuming one of the above four configurations of ranks can always
be brought, by subtracting properly chosen symmetric fully product vector,
to a PPT entangled state of ranks either $(5,7,7)$ or
$(5,7,8)$. Interestingly, with the half-analytical-half-numerical method
presented in Ref. \cite{Tura} as well as the numerical algorithm
described in Sec. \ref{sec:SymmExtr}, we found solely examples of
PPT entangled states of ranks $(5,7,8)$ (due to Ineq. (\ref{Extr_Ineq2})
PPT states of ranks $(5,7,9)$, $(5,8,8)$, and clearly $(5,8,9)$ cannot be extremal).
All the states of ranks $(5,7,7)$ found with the above algorithm were separable. This
together with the analytical considerations enclosed in appendix
\ref{app} suggest that that symmetric four-qubit states of these
ranks are generically separable. Provided this is the case, the analysis of
PPT entangled symmetric states of four-qubits could be reduced significantly
to the characterization of states with a single configuration of ranks $(5,7,8)$.

\textit{N=5.} In this case theorem \ref{thm_sep2} implies that five-qubit PPT
symmetric states are separable if either $r(\rho)\leq 5$ or
$r(\rho^{T_A})\leq 5$, or $r(\rho^{T_{AB}})\leq 4$ are separable.
Then, theorem \ref{sep_gen} says that if either $r(\rho^{T_A})\leq
8$ or $r(\rho^{T_{AB}})\leq 9$, they are
generically separable. Similarly to the case of $N=4$, this
leaves six (out of 120 possible under the assumption that $r(\rho)=6$)
ranks for which typical PPT symmetric states need not be separable:
$(6,9,10)$, $(6,9,11)$, $(6,9,12)$, $(6,10,10)$, $(6,10,11)$, and $(6,10,12)$.

With the aid of theorem \ref{thm:edge0}, one sees that five-qubit
PPT symmetric states of ranks $(6,9,12)$, $(6,10,11)$, and $(6,10,12)$ are
generically not edge (notice that due to Ineq. \ref{Extr_Ineq2}
for the same ranks PPT states cannot be extremal). Hence,
analysis of PPT entanglement in in this case reduces to three configurations of
ranks $(6,9,10)$, $(6,10,10)$, and $(6,9,11)$. Interestingly, only
in the second case we found examples of extremal states with the
above numerical algorithm (see table \ref{Excl_Ranks}).

{\it N=6.} Let us finally consider the case of six qubits.
Theorem \ref{thm_sep2} states that such PPT states are separable provided
that either $r(\rho)\leq 6$ or $r(\rho^{T_A})\leq 6$, or $r(\rho^{T_{AB}})\leq 5$, or
$r(\rho^{T_{ABC}})\leq 4$. Moreover, theorem \ref{sep_gen} implies that
they are generically separable if either $r(\rho^{T_A})\leq 10$ or
$r(\rho^{T_{AB}})\leq 12$, or $r(\rho^{T_{ABC}})\leq 12$.
The number of the remaining configurations of ranks among which
one may seek PPT entangled states is then 24 (out of all possible 2880 when assumed that
$r(\rho)=7$), which is considerably larger than the corresponding numbers for $N=4,5$.

Then, from theorem \ref{thm:edge1} it follows that six-qubit
PPT states of ranks $(7,12,15,15)$ are not edge and
theorem \ref{thm:edge0} says that generic PPT states of ranks
$(7,12,14,16)$ and $(7,11,15,16)$ are also not edge. This, together with
the fact that states of maximal ranks, i.e., $(7,12,15,16)$ are not edge,
allows to reduce the problem of characterization of six-qubit PPT states to
still quite large number of 20 configurations. There are, nevertheless,
only two sets of ranks for which, using the algorithm from Sec.
\ref{sec:algor}, we found extremal PPT entangled states.

\section{Conclusion}
\label{sec:concl}

Let us briefly summarize the obtained results. Our aim was to
characterize PPT entanglement in symmetric states. We have made a
significant step towards reaching this goal, yet the complete
characterization for the general case remains open.

First, we have derived simple separability criteria for PPT
symmetric states in terms of their ranks, complementing the criterion
stated in Ref. \cite{EckertAnP}. Interestingly, these criteria
imply that for most of the possible configurations of ranks, PPT
symmetric states are generically separable, and PPT entanglement
may appear only in a small fraction of cases, vanishing for large
number of parties. Putting $r(\rho)=N+1$, there are precisely $(\lfloor N/2\rfloor+1)!$
such configurations out of $N!(N/2+1)^{2\left(\lfloor N/2 \rfloor -N/2  \right)+1}$
all possible ones. For the exemplary cases
of four and five qubits this gives six different sets (out of,
respectively, 72 and 120 all possible ones) of ranks for which
typical PPT states need not be separable.

Second, we have singled out some of the configurations
of ranks for which PPT symmetric states are generically not edge,
allowing for further reduction of relevant configurations of ranks.
This is because if a PPT entangled state is not edge it
can be decomposed as a convex combination of a pure product vector
and a PPT symmetric state of lower ranks. From this point of view
the relevant configurations of ranks are those than cannot be
further reduced by subtracting a product vector from the state.
Again, in the particular case of small systems consisting of four
and five qubits, PPT states of higher ranks are generically not
edge, lowering the number of the configurations to treatable two
and three, respectively, for $N=4$ and $N=5$.

Finally, with the aid of the algorithm proposed in Ref.
\cite{Leinaas}, we have searched for extremal PPT symmetric
states. We have investigated systems consisting of 4 up to 23
qubits and encountered a clear pattern behind the configurations
of ranks for which we have found examples of extremal states. In
particular, for even $N$, except for the cases of $N=6$ and $N=8$,
there are always three configurations (following the same pattern)
of ranks. Interestingly, for odd $N$ there is only a single such
configuration, i.e., all ranks of the state and its partial
transpositions are maximal except for the last one (the partial
transposition with respect to half of the qubits), which amounts
to two less than maximal. This is somehow in contrary to the
intuition that the number of different sets of ranks for which one
finds extremal states should grow. On the other hand, it indicates
that the problem of characterization of PPT entanglement in
symmetric states could further be reduced to just few different
types of states.

\begin{acknowledgements}
Discussion with L. Chen, P. Hyllus, M. Ku\'s, and G. T\'oth are
acknowledged. This work is supported by EU project AQUTE, ERC
Grant QUAGATUA, EU projects CHIST-ERA DIQIP, Spanish MINCIN
project FIS2008-00784 (TOQATA). R. A. acknowledges the support
from the Spanish MINCIN through the Juan de la Cierva program. M.
L. acknowledged the Alexander von Humboldt Foundation. J. S.
acknowledges ICFO for kind hospitality.
\end{acknowledgements}

\appendix

\section{The case of $(5,7,7)$}
\label{app}

We consider here four-qubit PPT symmetric states of ranks
$(5,7,7)$ and provide a possible way of proving that generic
states of these ranks are separable. For this purpose we use the
approach developed in Ref. \cite{Normal}. First, recall that any
state $\rho$ acting on $\mathcal{H}$ can be written as a convex
combination of rank-one projectors
\begin{equation}\label{decomposition}
\rho=\sum_{k=1}^{l}\proj{\psi_k},
\end{equation}
where the unnormalized vectors $\ket{\psi_k}$ are in general
nonorthognal (a particular example of such decomposition is the
eigendecomposition of $\rho$).

Denoting by $\ket{e_k}$ orthonormal vectors spanning
$\mathcal{H}$, we see that any element of $\rho$ in this basis can
be written as
\begin{eqnarray}
\langle e_i|\rho|e_j\rangle=\sum_{k=1}^{l}\langle
e_i|\psi_k\rangle\!\langle\Psi_k|e_j\rangle =\langle v_i|v_j
\rangle
\end{eqnarray}
for any $i,j=1,\ldots,\dim\mathcal{H}$, where the
$l$-dimensional vectors $\ket{v_i}$ are defined as
\begin{equation}
\ket{v_i}=\left(
\begin{array}{c}
\langle\psi_1|e_i\rangle\\
\vdots \\
\langle\psi_l|e_i\rangle
\end{array}
\right)\qquad (i=1,\ldots,\dim\mathcal{H}).
\end{equation}
Consequently, $\rho$ is the so-called Gram matrix, i.e., the
matrix of scalar products of a set of vectors $\{\ket{v_i}\}$
called further the Gram system of $\rho$. A different
decomposition in Eq. (\ref{decomposition}) leads to a different
Gram system and all Gram systems of $\rho$ are related via unitary
matrices (if extended to a properly large Hilbert space).

Let us now consider a four-qubit PPT symmetric state $\rho$ of
ranks $(5,7,7)$ and find its Gram system together with the Gram systems of
$\rho^{T_A}$, $\rho^{T_B}$, and $\rho^{T_{AB}}$, starting from $\rho^{T_A}$.
Since, by assumption, the latter is positive and $r(\rho^{T_A})=7$,
it admits a decomposition as in Eq. (\ref{decomposition})
(for instance the eigendecomposition) with seven
rank-one components, i.e.,
\begin{equation}\label{decomposition2}
\rho^{T_A}=\sum_{k=1}^{7}\proj{\Psi_k},
\end{equation}
where the vectors $\ket{\Psi_k}$ are subnormalized.

Denoting by $\mathcal{B}$ the standard product basis
$\{\ket{i,j,k,l}\}$ of $(\mathbbm{C}^2)^{\ot 4}$ and utilizing the
fact that every $\ket{\Psi_k}$ in (\ref{decomposition}) belongs to
$\mathbbm{C}^2\ot\mathcal{S}_3$, one finds that the Gram system of
$\rho^{T_A}$ with respect to $\mathcal{B}$ consists of eight
seven-dimensional vectors $\ket{a},\ldots,\ket{d}$ and
$\ket{\widetilde{a}},\ldots,\ket{\widetilde{d}}$ whose elements
are given by
\begin{eqnarray}
a_k&=&\langle\Psi_k|0000\rangle,\nonumber\\
b_k&=&\langle\Psi_k|0001\rangle=\langle\Psi_k|0010\rangle=\langle\Psi_k|0100\rangle,\nonumber\\
c_k&=&\langle\Psi_k|0011\rangle=\langle\Psi_k|0101\rangle=\langle\Psi_k|0110\rangle,\nonumber\\
d_k&=&\langle\Psi_k|0111\rangle,\nonumber\\
\widetilde{a}_k&=&\langle\Psi_k|1000\rangle,\nonumber\\
\widetilde{b}_k&=&\langle\Psi_k|1001\rangle=\langle\Psi_k|1010\rangle=\langle\Psi_k|1100\rangle,\nonumber\\
\widetilde{c}_k&=&\langle\Psi_k|1011\rangle=\langle\Psi_k|1101\rangle=\langle\Psi_k|1110\rangle,\nonumber\\
\widetilde{d}_k&=&\langle\Psi_k|1111\rangle.
\end{eqnarray}
Let us then introduce the following $7\times 4$ matrices
$A=(\ket{a},\ket{b},\ket{b},\ket{c})$,
$B=(\ket{b},\ket{c},\ket{c},\ket{d})$,
$\widetilde{B}=(\ket{\widetilde{a}},\ket{\widetilde{b}},\ket{\widetilde{b}},\ket{\widetilde{c}})$,
and
$\widetilde{C}=(\ket{\widetilde{b}},\ket{\widetilde{c}},\ket{\widetilde{c}},\ket{\widetilde{d}})$,
with columns given by the vectors
$\ket{a},\ldots,\ket{\widetilde{d}}$. Then $\rho^{T_A}$ can be
written as
\begin{eqnarray}\label{matrix}
\rho^{T_A}&=&\left(
\begin{array}{cccc}
A^{\dagger}A & A^{\dagger}B & A^{\dagger}\widetilde{B} & A^{\dagger}\widetilde{C}\\
B^{\dagger}A & B^{\dagger}B & B^{\dagger}\widetilde{B} & B^{\dagger}\widetilde{C}\\
\widetilde{B}^{\dagger}A & \widetilde{B}^{\dagger}B & \widetilde{B}^{\dagger}\widetilde{B} & \widetilde{B}^{\dagger}\widetilde{C}\\
\widetilde{C}^{\dagger}A & \widetilde{C}^{\dagger}B & \widetilde{C}^{\dagger}\widetilde{B} & \widetilde{C}^{\dagger}\widetilde{C}\\
\end{array}
\right)\nonumber\\&=&(A^{\dagger}\,\, B^{\dagger}\,\,
\widetilde{B}^{\dagger}\,\,
\widetilde{C}^{\dagger})\left(\begin{array}{c}A\\B\\\widetilde{B}\\\widetilde{C}
\end{array}\right),
\end{eqnarray}
where $X^{\dagger}Y$ $(X,Y=A,\widetilde{B},\widetilde{C})$ denotes a $4\times 4$ matrix
consisting of scalar products of vectors defining $X$ and $Y$. We
will then symbolically denote $\rho^{T_A}=(A\,\, B\,\,
\widetilde{B}\,\, \widetilde{C})$.

In the same way we can represent $\rho$. Since it is symmetric it
admits the form $\rho=(A'\,\,B'\,\,B'\,\,C)$ with $A',B'$, and $C$
constructed in the same way as $A$, $B$, etc. from the Gram system
of $\rho$. Now, since both three-qubit matrices $\langle
0|\rho|0\rangle$ and $\langle 0|\rho^{T_A}|0\rangle$ arising by
projecting the first qubit of $\rho$ and $\rho^{T_A}$ onto
$\ket{0}$, are equal, they have the same Gram systems and
therefore there is a unitary $U$ such that $A'=UA$ and $B'=UB$.
Then, since by multiplying by a unitary operator all the vectors
of the Gramm system of $\rho$ one gets another Gram system, we can
always set $A'=A$ and $B'=B$.

Analogously, one sees that the matrices $\langle 1|\rho|1\rangle=\langle
1|\rho^{T_A}|1\rangle$ (arising by projecting the first qubit onto $\ket{1}$),
and therefore there is a unitary $U$ such
that $\widetilde{B}=UB$ and $\widetilde{C}=UC$. In conclusion, we
see that $\rho$ and $\rho^{T_A}$ can be represented as
\begin{equation}
\rho=(A\,\,B\,\,B\,\,C),\qquad
\rho^{T_A}=(A\;\;B\;\;U\!B\;\;U\!C).
\end{equation}
By taking partial transposition of $\rho$ with respect to $A$ and
comparing it with the above representation of $\rho^T_A$, one gets
some conditions for $U$:
\begin{equation}\label{cond1}
\begin{array}{cc}
A^{\dagger}UB=B^{\dagger}A, \quad&\quad A^{\dagger}UC=B^{\dagger}B,\\
B^{\dagger}UB=C^{\dagger}A,\quad &\quad B^{\dagger}UC=C^{\dagger}B.\\
\end{array}
\end{equation}

The same reasoning allows us to represent
$\rho^{T_B}$ and $\rho^{T_{AB}}$ in the following way
\begin{eqnarray}\label{Gramm2}
\rho^{T_B}&=&(A\;\; U\!B\;\; B\;\; U\!C),\nonumber\\
\rho^{T_{AB}}&=&(A\;\;U\!B\;\;V\!B\;\;V\!U\!C)
\end{eqnarray}
with $V$ being a unitary matrix and $UB=VB$. Again, comparison of
$\rho^{T_{AB}}$ given by Eq. (\ref{Gramm2}) to the partial
transposition of $\rho^{T_B}$ with respect to $A$, gives further
conditions
\begin{equation}\label{cond2}
\begin{array}{cc}
A^{\dagger}VB=B^{\dagger}A,\quad & \quad A^{\dagger}VUC=B^{\dagger }UB,\\
B^{\dagger}U^{\dagger}VB=C^{\dagger}U^{\dagger}A,\quad &\quad
B^{\dagger}U^{\dagger}VUC=C^{\dagger}B.
\end{array}
\end{equation}

Now, having Gram systems of $\rho$ and its partial transpositions,
we introduce the following matrix
\begin{equation}\label{defQ}
Q=\sum_{k=1}^{7}[\ket{0}\ket{\Psi_k}+\ket{1}\ket{\Phi_k}]
\end{equation}
with $\ket{\Psi_k}$ and $\ket{\Phi_k}$ denoting decompositions of
$\rho^{T_B}$ and $\rho^{T_{AB}}$ [as in (\ref{decomposition2})]
corresponding to their Gram systems introduced above, and
$[\ket{\psi}]$ standing for a projection onto $\ket{\psi}$. In
terms of the Gram systems (\ref{Gramm2}), $Q$ assumes the form
\begin{equation}
Q=(A\;\;U\!B\;\;B\;\;U\!C\;\;A\;\;U\!B\;\;V\!B\;\;V\!U\!C).
\end{equation}
By careful counting of the dimensions, one notices that $Q$ is an
unnormalized density matrix acting on
$\mathbbm{C}^7\ot\mathbbm{C}^3$ with respect to the bipartition
$aAB|CD$, where $a$ denotes the auxiliary subsystem [cf.
(\ref{defQ})], while $ABCD$ stand for the subsystems of $\rho$. Also,
by definition $Q$ is of rank seven, and hence according to Ref.
\cite{MN} $Q$ is separable with respect to this bipartition
provided that it is supported on $\mathbbm{C}^7\ot\mathbbm{C}^3$ and
$Q^{T_{aAB}}\geq 0$. Although we cannot prove the former condition,
it is clear that generic $Q$ is supported on $\mathbbm{C}^7\ot\mathbbm{C}^3$.
Then, after direct algebra and with
the aid of conditions (\ref{cond1}) and (\ref{cond2}), one sees
that the latter condition, i.e., $Q^{T_{aAB}}\geq 0$
holds if the following two equations
\begin{equation}
B^{\dagger}VUC=C^{\dagger}UB,\qquad
C^{\dagger}U^{\dagger}VUC=C^{\dagger}UC
\end{equation}
are obeyed. Still, exploiting the explicit forms of
$B$ and $C$, the above conditions can be further simplified, leading to
a set of equations for scalar product of vectors composing the Gram system of $\rho$.
Some of them, by virtue of the Eqs. (\ref{cond1}) and (\ref{cond2})
are immediately satisfied. Provided that the remaining equations also hold,
one has that indeed $Q^{T_{aAB}}\geq 0$ and generic $Q$ is separable.

Then, it is clear that by projecting the auxiliary qubit $a$ of
$Q$ onto $\ket{0}$, one recovers $\rho$. So, if $Q$ is separable
across $aAB|CD$, then $\rho$ is separable across $AB|CD$, implying
that, due to the underlying symmetry, it is fully separable.

\end{document}